\documentclass[a4paper,12pt]{article}
\usepackage{soul}
\usepackage[usenames,dvipsnames]{color}
\usepackage{jheppub}
\usepackage{esvect}
\usepackage{amsmath, amssymb, slashed, epsf, color, graphicx, latexsym, bm}
\usepackage{tensor}
\usepackage{physics}
\usepackage{epsfig}
\usepackage{graphics}
\usepackage{mathtools}
\usepackage{caption}
\usepackage{float}
\usepackage{bbm}
\usepackage{subcaption}
\usepackage{enumitem}
\usepackage{verbatim}
\usepackage{blindtext}
\usepackage[mathlines]{lineno}

\begin{document}

\title{Asymptotic Symmetry of Four Dimensional Einstein-Yang-Mills and Einstein-Maxwell Theory}

\author[a]{Nabamita Banerjee,}
\author[a]{Tabasum Rahnuma,}
\author[b]{and Ranveer Kumar Singh.}
\affiliation[a]{Indian Institute of Science Education and Research Bhopal,
	Bhopal Bypass, Bhopal 462066, India.}
\affiliation[b]{NHETC and Department of Physics and Astronomy, Rutgers University, 126
Frelinghuysen Rd., Piscataway NJ 08855, USA}

\emailAdd{nabamita@iiserb.ac.in}
\emailAdd{tabasum19@iiserb.ac.in}
\emailAdd{ranveersfl@gmail.com}

\abstract{Asymptotic symmetry plays an important role in determining physical observables of a theory. Recently, in the context of four dimensional asymptotically flat pure gravity and $\mathcal{N}=1$ supergravity, it has been proposed that OPEs of appropriate celestial amplitudes can be used to find their asymptotic symmetries. In this paper we find the asymptotic symmetry algebras of four dimensional Einstein-Yang-Mills and Einstein-Maxwell theories using this alternative approach, namely using the OPEs of their respective celestial amplitudes. The algebra obtained here are in agreement with the known results in the literature.}

\maketitle

\section{Introduction and Summary}
    Both gauge and gravity theories in 4 dimensional asymptotically flat spacetime have infinite number of non-trivial asymptotic symmetries. For these theories, the soft theorems, infinite dimensional asymptotic symmetries and the memory effects are the three different manifestations of a single framework. This is wonderfully portrayed in  Strominger's Infrared triangle \cite{Strominger:2017zoo}. In the context of gravity, the infinite dimensional asymptotic symmetry is called the BMS symmetry. The passage of a gravitational radiation pulse through a nearby detector induces a relative displacement in the detector position. This effect is measurable and is known as gravitational memory effect\footnote{The relation between memory and asymptotic symmetries is given by a universal formula which also has been extended for gauge theories \cite{Pasterski:2015zua}.}.  On the other hand, due to the passage of a gravitational radiation, the difference in the initial and final geometries of the spacetime is related by a BMS supertranslation \cite{Strominger:2014pwa}. The Ward identities of BMS supertranslation invariance in quantum gravity are expressed as data representing gravitational radiation at the null infinity \cite{Strominger:2013jfa}. The third point of the triangle is related to the soft theorem. Amplitudes involving one or multiple soft graviton\footnote{A graviton is called soft when all components of its momentum go to a vanishing limit.} currents (in terms of momentum space creation and annihilation operators) are given by Weinberg's soft graviton theorem and the supertranslaion Ward identity can be reproduced from this soft theorem \cite{He:2014laa}. This is how the infrared triangle is formed. However, quite independently the relation between the memory and soft theorems has been established  by Strominger \textit{et. al.} \cite{Strominger:2017zoo, Strominger:2014pwa}.

    The usual method of finding asymptotic symmetry of a theory is governed by finding symmetry transformation parameters for various fields that preserve their falloff conditions at the boundary. For a theory of gravity, we look for asymptotic isometry transformations that leave the boundary falloffs of various gravitational fields intact. The falloff conditions are determined with respect to the asymptotic geometry. The BMS symmetry group of gravity is known after Bondi-Metzner-Sachs who first studied the asymptotic symmetries of Einstein gravity in four spacetime dimensions \cite{Bondi:1962px, Sachs:1962wk, Sachs:1962zza}. The Lie algebra $\mathfrak{bms_4}$ is defined as the semi-direct sum of the Lie algebra of conformal killing vectors of the Riemann(celestial) sphere corresponding to infinitesimal local conformal transformations with supertranslation generators on the Riemann sphere \cite{Barnich:2009se}. Barnich \textit{et. al.} provided the detailed derivation of the generalisaion of $\mathfrak{bms_4}$ algebra\footnote{and $\mathfrak{bms_3}$} along with a proof that the asymptotic symmetry algebra in four dimension is represented by spacetime vectors which can be generalised to the extended algebra via conformal rescaling of the boundary metric \cite{Barnich:2010eb}. The surface charges associated with asymptotic symmetries in $4D$ flat spacetimes at null infinity and their transformation properties are constructed. The divergence and non-divergence of the supertranslation and superrotation charges in case of BMS algebra and Kerr black hole have been studied \cite{Barnich:2011mi}.
    
 Asymptotic symmetry analysis via killing vectors is often tedious and challenging. So far this prescription has only been used for obtaining the asymptotic symmetry groups of pure gravity theory. In three space time dimensions, the alternative Chern-Simons formulation of (super)gravity\footnote{asymptotic symmetries for various 3 dimensional supergravity theories can be found in \cite{Barnich:2014cwa, Banerjee:2016nio, Lodato:2016alv,Banerjee:2017gzj, Fuentealba:2017fck, Banerjee:2018hbl, Banerjee:2019lrv} and for higherspin-gravity can be found in \cite{Gonzalez:2013oaa, Afshar:2013vka, Banerjee:2015kcx}.} has turned out to be the most useful tool for obtaining the asymptotic symmetric algebra. But for four spacetime dimensions that we are currently interested in, gravity does not have a Chern-Simons formulation. Thus an alternate method for computing the asymptotic symmetry algebra for a theory of gravity in four dimensions is desirable, so that the enhanced symmetry group in presence of supersymmetry and other internal symmetries can be obtained. The recent uncovering of the relation between the on-shell physics of asymptotically flat theories and 2D CFT has proved to be a powerful tool in the computation of BMS algebra. Let us briefly discuss the relation here : \\
 \begin{itemize}
 \item
 It is a well-known fact that the bulk symmetry $\mathrm{SL}(2, \mathbb{C})$ of an asymptotically flat four dimensional theory is identical to the global part of a two dimensional Conformal Field Theory (CFT). Given fields in the bulk of an asymptotically flat theory, one can associate conformal operators with these fields that live on the two dimensional sphere, namely the celestial sphere denoted by $\mathcal{CS}^2$ sitting at the null boundaries of the space time. 
 \item The boundary physics is captured by a 2D CFT known as  celestial CFT (CCFT) of these operators on $\mathcal{CS}^2$. In this \textit{celestial} approach the non-trivial symmetries and their algebra of the bulk theory can be computed using the $2$D conformal invariance of the CCFT correlators. In particular the four dimensional scattering amplitudes of the bulk theory is related via Mellin transformation to the conformal correlators of the CCFT operators. Such CCFT correlators, that are associated with bulk scattering amplitudes, are called celestial amplitudes. 
 \item We then construct conserved currents of the CCFT via shadow transform \cite{Pasterski:2017kqt} of the conformal operators and compute the OPEs of various currents. The OPEs give us the algebra of the Laurent modes of the currents using standard methods of 2D CFT. After quantization these modes act as the generators of the BMS algebra. 
 \end{itemize}
 Thus the computations of asymptotic symmetry algebra reduces to the computation of appropriate OPEs which in turn depends on the soft and collinear limits in the bulk. We discuss these two limits in details in the main draft.
    
     Celestial amplitudes are of immense importance as an independent entity in itself \cite{Schreiber:2017jsr}. As stated earlier, there are infinite number of symmetries in asymptotically flat theories. These symmetries impose constraints on the celestial amplitudes via Ward identities which give us the soft theorems in terms of celestial amplitudes \cite{Fan:2019emx}.  Since the bulk scattering amplitudes are related to celestial amplitudes via Mellin transformation, these constraints in turn imply constraints on the bulk scattering amplitude. Thus a more thorough understanding of celestial amplitudes can uncover new symmetries and constraints on the bulk amplitudes.
    
    In the present work we have computed the asymptotic BMS algebra in Einstein-Yang-Mills (EYM) and Einstein-Maxwell (EM) theories using the above discussed CCFT technique. Both the symmetry algebras are already known in the literature \cite{Barnich:2013sxa, Henneaux:2018hdj} and our results match with them. Thus it provides a check on the alternate prescription of finding asymptotic symmetry algebras for four dimensional flat theories. The paper is organized as follows. In Section \ref{Notation} we record the preliminaries for writing the currents including the properties of conformal primaries and set up notations for the rest of the paper. In Section \ref{AspEYM}, we define the current corresponding to the conformal gauge Boson operator in EYM theory and write the OPE with spin 1 primary operators. In Section \ref{opeEYMBMS}, we derive the OPEs of different combination of the EYM currents with the supertranslation and superrotation generators of pure BMS symmetry. We further construct a composite current and elaborate on its significance. 
    In Section \ref{bmsEYM}, we compute the symmetry algebra from the modes of currents and find the $\mathrm{u}(N)$ extended $\mathfrak{bms_4}$ algebra. In Section \ref{asyEM}, we define the current corresponding to the conformal operator in a $\mathrm{u}(1)$ gauge theory and use the results of previous sections with appropriate changes to compute the asymptotic symmetry algebra of EM theory. We end the paper with discussions and open questions. 
    Finally, in the Appendix, we compute certain integrals using the global conformal invariance of the celestial correlators and verify the conformal dimension of our normalized current using the traditional method of constructing OPE with the superrotation generators. 
    \section{Notations and Preliminaries} \label{Notation}

By now it is a well known fact that the scattering amplitudes of a four dimensional theory in Minkowski spacetime can be cast into appropriate quantities on the celestial sphere via certain mappings of the momenta and amplitudes \cite{Pasterski:2016qvg, Pasterski:2017kqt, Pasterski:2017ylz, Schreiber:2017jsr}. The four dimensional Lorentz group $\mathrm{SL}(2,\mathbb{C})$ is equivalent to the global part of two dimensional conformal group and acts on points of the celestial sphere $\mathcal{CS}^2$ via fractional linear transformation. To elaborate on this connection, for four dimensional spacetime we use Bondi-coordinates $(u,r,z,\bar{z})$ where $(z,\bar{z})$ parametrise the celestial sphere at null infinity. Then $\mathrm{SL}(2,\mathbb{C})$  acts on  $\mathcal{CS}^2$ as follows:
\[
(z,\bar{z})\longmapsto\left(\frac{az+b}{cz+d},\frac{\bar{a}\bar{z}+\bar{b}}{\bar{c}\bar{z}+\bar{d}}\right),\quad \begin{pmatrix}
a&b\\c&d
\end{pmatrix}\in\mathrm{SL}(2,\mathbb{C}).
\]
A general null momentum vector $p^{\mu} : p^2=0$ can be parametrized as 
\[
p^{\mu}=\omega q^{\mu},\quad q^{\mu}=\frac{1}{2}\left(1+|z|^{2}, z+\bar{z},-i(z-\bar{z}), 1-|z|^{2}\right),
\]
where $q^{\mu}$ is a null vector, and $\omega$ is identified with the light cone energy. Under the Lorentz group the four momentum transforms as a Lorentz vector $p^{\mu}\mapsto \Lambda^{\mu}_{~\nu}p^{\nu}$. This induces the following transformation of $\omega$ and $q^{\mu}$:
$$
\omega \mapsto (c z+d)(\bar{c} \bar{z}+\bar{d}) \omega, \quad q^{\mu} \mapsto q^{\prime \mu}=(c z+d)^{-1}(\bar{c} \bar{z}+\bar{d})^{-1} \Lambda_{~\nu}^{\mu} q^{\nu}.
$$
The wave functions in the four dimensional Minkowski space is mapped to particular operators on the celestial sphere via Mellin transforms. They are also called conformal primary wavefunctions. This is done as follows: the plane wave packets corresponding to gauge boson and graviton is given as \cite{Fan:2019emx, Fotopoulos:2019tpe,  Fotopoulos:2019vac},
\[
\epsilon_{\mu}^{\ell}(p) e^{\mp i\left|p_{0}\right| X^{0} \pm i \vec{p} \cdot \vec{X}} \quad \text{and}\quad\epsilon_{\mu\nu}^{\ell}(p) e^{\mp i\left|p_{0}\right| X^{0} \pm i \vec{p} \cdot \vec{X}}
\]
respectively where $\ell$ is the helicity and $\epsilon_{\mu}^{\ell}$ and $\epsilon_{\mu\nu}^{\ell}$ represents the polarisations of the spin 1 ans spin 2 particles respectively.
Their Mellin transforms give the
corresponding operators on $\mathcal{CS}^{2}:$
$$
V_{\mu}^{\Delta, \ell}\left(X^{\mu}, z, \bar{z}\right) \equiv \partial_{J} q_{\mu} \int_{j}^{\infty} d \omega\; \omega^{\Delta-1} e^{\mp i \omega q\cdot X-\epsilon \omega} \quad(\ell=\pm 1)
$$
$$
H_{\mu \nu}^{\Delta, \ell}\left(X^{\mu}, z, \bar{z}\right) \equiv \partial_{J} q_{\mu} \; \partial_{J} q_{\nu} \int_{0}^{\infty} d \omega \; \omega^{\Delta-1} e^{\mp i \omega q\cdot X-\epsilon \omega}, \quad(\ell=\pm 2)$$\\
where, $\partial_{J}=\partial_z$ for $\ell=+1,+2$ , $\partial_{J}=\partial_{\bar{z}}$ for
$\ell=-1,-2$ 
and $\Delta$ is the ``conformal dimension" of the conformal primary wavefunctions corresponding to gauge bosons and graviton. It belongs to the principal continuous series of irreducible unitary $\mathrm{SO}(1,3)$ representation \cite{Pasterski:2017kqt}, $\Delta \in 1+i \mathbb{R}$. These conformal dimensions are the Mellin-dual to the energies.
The conformal wave functions corresponding to the graviton and gauge boson, upto gauge and diffeo transformations respectively, can be then
written as:
\begin{equation}
    \begin{split}
&A_{\mu}^{\Delta, \ell}=g(\Delta) V_{\mu J}^{\Delta, \ell}+\text{gauge}\\
&G_{\mu \nu}^{\Delta, \ell}=f(\Delta) H_{\mu \nu}^{\Delta, \ell}+\text{diffeo}
\end{split}
\label{eq:conwavfugragauge}
\end{equation}
where 
\begin{equation}
    g(\Delta)=\frac{\Delta-1}{\Gamma(\Delta+1)}, \quad f(\Delta)=\frac{1}{2} \frac{\Delta(\Delta-1)}{\Gamma(\Delta+2)}
\label{eq:normconstconfop}    
\end{equation}
are the normalisation constants\footnote{The presence of normalisation constants ($g(\Delta)$,  $f(\Delta)$) fixes the fields with spin 1 and spin 2 to be pure gauge and pure diffeomorphisms respectively under soft \textit{conformal} limits \cite{Fotopoulos:2019vac}, $\Delta \to 1$ and $\Delta \to 0,1$. These factors implement the CPT symmetry of the 4D theory at the level of celestial CFT \cite{Fan:2019emx}. }.
The conformal correlators are related to the scattering amplitudes in the conformal basis via Mellin transformations:
\begin{equation}
    \begin{aligned}
\left\langle\prod_{n=1}^{N} \mathcal{O}_{\Delta_{n}, \ell_{n}}\left(z_{n}, \bar{z}_{n}\right)\right\rangle=\left(\prod_{n=1}^{N} c_{n}\left(\Delta_{n}\right) \int d \omega_{n} \omega_{n}^{\Delta_{n}-1}\right) \delta^{(4)}\left(\sum_{n=1}^{N} \epsilon_{n} \omega_{n} q_{n}\right)
\mathcal{M}_{\ell_{1} \ldots \ell_{N}}\left(\omega_{n}, z_{n}, \bar{z}_{n}\right)
\end{aligned}
\label{eq:Ocorscatamprel}
\end{equation}
where $\mathcal{M}_{\ell_{1} \ldots \ell_{N}}$ are EYM Feynman's scattering matrix elements with helicities $\ell_{n}$, $\epsilon_{n}=+1$ or $-1$ depending on whether the particles are incoming or outgoing respectively and $c_{n}$ are the normalization constants
$$
c_{n}\left(\Delta_{n}\right)=\left\{\begin{array}{ll}
g\left(\Delta_{n}\right) & \text { for } \ell_{n}=\pm 1, \\
f\left(\Delta_{n}\right) & \text { for } \ell_{n}=\pm 2,
\end{array}\right.
$$
where $f(\Delta)$ and $g(\Delta)$ are defined in equation \eqref{eq:normconstconfop}. We have suppressed the gauge indices for simplicity.

In the language of celestial CFT, the operator product expansion (OPE) of the conformal operators can be extracted from the correlator in (\ref{eq:Ocorscatamprel}) by taking the limits of coinciding insertion points on $\mathcal{CS}^2$.The same implies a collinear limit of momenta in the scattering amplitudes. The form of the OPEs depends on different gauge/gravity conformal operators, in various helicity combinations in their collinear limit.

In our context, when two massless particles propagate with parallel four-momenta they are said to be collinear to each other. The propagator has a pole when a virtual particle splits into a collinear pair via a three-point interaction. Such singularities can be separated out via \textit{collinear} limit \cite{Taylor:2017sph} where the only combined vertex contribute to the divergence. This way we can factorize our tree amplitudes.

As an example, in Yang-Mills (YM) theory the tree level amplitudes with two collinear momenta are given by the collinear poles \cite{Fan:2019emx}. Let the combined momentum of the collinear gluons be
$$
P^{\mu}=p_{n-1}^{\mu}+p_{n}^{\mu}=\omega_{P} q_{P}^{\mu},
$$
where the parametrizations are as follows
$$
\omega_{P}=\omega_{n-1}+\omega_{n}, \quad q_{P}^{\mu}=q_{n-1}^{\mu}=q_{n}^{\mu}.$$\\
At tree level, the partial amplitudes \footnote{Partial amplitudes are the color stripped amplitudes corresponding to a particular choice of the Chan-Paton factor Tr($T^1T^2\dots T^n$). See \cite{Taylor:2017sph} for details.} for possible helicity combinations are given by \cite{Fan:2019emx},
\begin{equation}
\begin{split}
\mathcal{M}\left(1, \ldots, n-1^{+}, n^{+}\right) &=\frac{1}{z_{(n-1) n}} \frac{\omega_{P}}{\omega_{n-1} \omega_{n}} \mathcal{M}\left(1, \ldots,n-2, P^{+}\right)+\ldots \\
\mathcal{M}\left(1, \ldots, n-1^{-}, n^{-}\right) &=\frac{1}{\bar{z}_{(n-1) n}} \frac{\omega_{P}}{\omega_{n-1} \omega_{n}} \mathcal{M}\left(1, \ldots, n-2,P^{-}\right)+\ldots\\
\mathcal{M}\left(1, \ldots, n-1^{-}, n^{+}\right)&= \frac{1}{z_{(n-1) n}} \frac{\omega_{n-1}}{\omega_{n} \omega_{P}} \mathcal{M}\left(1, \ldots, n-2, P^{-}\right)\\
& \hspace{6ex}+\frac{1}{\bar{z}_{(n-1) n}} \frac{\omega_{n}}{\omega_{n-1} \omega_{P}} \mathcal{M}\left(1, \ldots,n-2, P^{+}\right)+\ldots.
\end{split}
\end{equation}
where $z_{(n-1) n}= z_n - z_{(n-1)}.$ Let us explain the notation used in the above expression. 
In LHS, we have the partial amplitudes $\mathcal{M}\left(1, \ldots, (n-1)^{\alpha}, n^{\beta}\right)$ of $n$ gauge bosons with two adjacent gauge bosons $n-1$, $n$ with their specific helicities $\alpha, \beta = \pm 1$. In the RHS we have $n-1$ point partial gauge boson amplitude with the $(n-1)$-th gauge boson having the combined momenta $P$ of the collinear pair with helicity $\alpha=\pm 1.$ The RHS also contains leading collinear poles corresponding to the adjacent bosons. After Mellin transforming these amplitudes as in (\ref{eq:Ocorscatamprel}) we extract all the OPEs corresponding to the partial amplitudes as follows,
\begin{equation}
    \begin{split}
\mathcal{O}_{\lambda_{1}+}^{a}(z, \bar{z}) \mathcal{O}_{\lambda_{2}+}^{b}(w, \bar{w})&=\frac{C_1}{z-w} \sum_{c} f^{a b c} \mathcal{O}_{\left(\lambda_{1}+\lambda_{2}\right)+}^{c}(w, \bar{w})+\ldots\\
 \mathcal{O}_{\lambda_{1}-}^{a}(z, \bar{z}) \mathcal{O}_{\lambda_{2}-}^{b}(w, \bar{w})&=\frac{C_{2}}{\bar{z}-\bar{w}} \sum_{c} f^{a b c} \mathcal{O}_{\left(\lambda_{1}+\lambda_{2}\right)-}^{c}(w, \bar{w}) \ldots\\
\mathcal{O}_{\lambda_{1}-}^{a}(z, \bar{z}) \mathcal{O}_{\lambda_{2}+}^{b}(w, \bar{w})&= \frac{C_{3}}{z-w} \sum_{c} f^{a b c} \mathcal{O}_{\left(\lambda_{1}+\lambda_{2}\right)-}^{c}(w, \bar{w}) +\frac{C_4}{\bar{z}-\bar{w}} \sum_{c} f^{a b c} \mathcal{O}_{\left(\lambda_{1}+\lambda_{2}\right)+}^{c}(w, \bar{w})+\ldots
    \end{split}
\end{equation}
where $\Delta_i=1+i\lambda_i$ and $C_i$ are the normalization constants given in \cite{Fan:2019emx}.\\

Likewise in a scattering process, when a virtual particle goes on shell we get a soft particle which results in soft singularities. In CCFT we assign a conformal operator to each of the soft particles whose soft energy ($\omega \to 0$) limit corresponds to conformal soft limit $\Delta \to 1$ in case of gauge Bosons and $\Delta \to 0$ and $\Delta \to 1$ in case of gravitons \cite{Fotopoulos:2019vac}. An $n-$point scattering amplitude in YM theory (in celestial basis) after Mellin transformation is given by,
\begin{equation}
    \begin{split}
        \mathcal{A}_{J_{1} \ldots J_{n}}\left(\Delta_{i}, z_{i}, \bar{z}_{i}\right)=\left(\prod_{i=1}^{n} g\left(\Delta_{i}\right) \int d \omega_{i}\; \omega_{i}^{\Delta_i -1}\right) \delta^{(4)}\left(\sum_{i} \epsilon_{i} \omega_{i} q_{i}\right) \mathcal{M}_{\ell_{1} \ldots \ell_{n}}\left(\omega_{i}, z_{i}, \bar{z}_{i}\right)
    \end{split}
\end{equation}
where the arguments $(\Delta_{i}, z_{i}, \bar{z}_{i})$ in both sides span over $n$ values.

As an example in YM theory, for any helicity configurations, after taking the soft limit of the $n$th particle, we have our celestial YM amplitude in Mellin space \cite{Fan:2019emx} as,
\begin{equation}
    \begin{split}
        \mathcal{A}_{J_{1}, J_{2}, \ldots, J_{n-1}\left(J_{n}=+1\right)}=(-i)\left(\frac{1}{z_{(n-1) n}}+\frac{1}{z_{n 1}}\right) \mathcal{A}_{J_{1}, J_{2}, \ldots, J_{n-1}}.
    \end{split}
\end{equation}
For negetive helicities, we have a similar anti-holomorphic relation.

Using the relation between the correlator and the scattering amplitude defined in Eq.(\ref{eq:Ocorscatamprel}), one can now write the correlator corresponding to the above amplitude by summing over all Chan-Paton factors \cite{Fan:2019emx} (see footnote 5). This gives  
\[
\begin{aligned}
&\left\langle\mathcal{O}_{\Delta_{1}, J_{1}}^{a_{1}} \mathcal{O}_{\Delta_{2}, J_{2}}^{a_{2}} \ldots \mathcal{O}_{\Delta_{n-1}, J_{n-1}}^{a_{n-1}} \mathcal{O}_{\Delta_{n}, J_{n}}^{a_{n}}\right\rangle
=\sum_{\sigma \in S_{n-1}} \mathcal{A}_{J_{1} J_{2} \ldots J_{n-1} J_{n}}^{\sigma} \operatorname{Tr}\left(T^{a_{1}} T^{a_{\sigma(2)}} \ldots T^{a_{\sigma(n-1)}} T^{a_{\sigma(n)}}\right).
\end{aligned}
\]
This gives us the Ward identity with the soft current,  $j^a(z)=\lim\limits_{\Delta\to 1}\mathcal{O}_{\Delta,+}^a(z,\bar{z})$,
\begin{equation}
\begin{split}
    &\left\langle j^{a}(z) \mathcal{O}_{\Delta_{1}, J_{1}}^{b_{1}}\left(z_{1}, \bar{z}_{1}\right) \mathcal{O}_{\Delta_{2}, J_{2}}^{b_{2}}\left(z_{2}, \bar{z}_{2}\right) \ldots \mathcal{O}_{\Delta_{n}, J_{n}}^{b_{n}}\left(z_{n}, \bar{z}_{n}\right)\right\rangle \\
&\quad=\sum_{i=1}^{n} \sum_{c} \frac{f^{a b_{i} c}}{z-z_{i}}\left\langle\mathcal{O}_{\Delta_{1}, J_{1}}^{b_{1}}\left(z_{1}, \bar{z}_{1}\right) \ldots \mathcal{O}_{\Delta_{i}, J_{i}}^{c}\left(z_{i}, \bar{z}_{i}\right) \ldots \mathcal{O}_{\Delta_{n}, J_{n}}^{b_{n}}\left(z_{n}, \bar{z}_{n}\right)\right\rangle.
\end{split}
\end{equation}
Similarly we can find the Ward identity for the anti-holomorphic current $\overline{j}^{a}(\bar{z})=\lim\limits_{\Delta\to 1}\mathcal{O}_{\Delta,-}^a(z,\bar{z})$. These Ward identities can also be derived using OPEs. In CCFT technique, the soft and collinear limits of the celestial amplitudes in a given theory are the necessary ingredients for asymptotic symmetry analysis. 
\section{Asymptotic Symmetry Generators in EYM Theory} \label{AspEYM}
\noindent
In this section, we look for operators in the corresponding CCFT of EYM theory that will generate the asymptotic symmetry of the theory at null infinity. As per our expectation in this case the asymptotic symmetry algebra will be an extension of the usual $\mathfrak{bms}_4$ algebra by an $\mathrm{u}(N)$ current. 
\subsection{The symmetry currents}
We first construct the currents in our theory. Following \cite{Fotopoulos:2019vac}, we define the energy momentum tensor
$T(z)$ (and $\overline{T}(\bar{z})$) as the shadow transform of
the $\Delta=0$ graviton conformal operator $\mathcal{O}_{0,-2}$  (and $\mathcal{O}_{0,+2}$):
\begin{equation}
\begin{aligned}
&T(z)=\frac{3 !}{2 \pi} \int d^{2} z^{\prime} \frac{1}{\left(z-z^{\prime}\right)^{4}} \mathcal{O}_{0,-2}\left(z^{\prime}, \bar{z}^{\prime}\right) \\
&\overline{T}(\bar{z})=\frac{3 !}{2 \pi} \int d^{2} z^{\prime} \frac{1}{\left(\bar{z}-\bar{z}^{\prime}\right)^{4}} \mathcal{O}_{0,+2}\left(z^{\prime}, \bar{z}^{\prime}\right)
\end{aligned}
\label{eq:TTbardef}
\end{equation}
The usual superrotations of $\mathfrak{bms}_4$ are generated by $T$ and $\overline{T}$. The supertranslations are generated by the supertranslation current $\mathcal{P}(z,\bar{z})$ that is constructed form graviton conformal operators with $\Delta=1$ as in \cite{Fotopoulos:2019vac}. For this purpose we first look for the currents $P(z)$,  $\overline{P}(z)$ which are defined as the level one descendant of the $\Delta=1$ graviton conformal operator as,
\begin{equation}
    \begin{aligned}
&P(z)=\partial_{\bar{z}} \mathcal{O}_{1,+2}(z, \bar{z}) \\
&\overline{P}(\bar{z})=\partial_{z} \mathcal{O}_{1,-2}(z, \bar{z}) .
\end{aligned}
\label{eq:PPbardef}
\end{equation}

Now the holomorphic and antiholomorphic currents can be written as \cite{Fotopoulos:2019vac},
\begin{equation}
    \begin{split}
        P(z)=\frac{1}{8 \pi i} \oint d \bar{z} \; \mathcal{P}(z, \bar{z}), \qquad \qquad  \overline{P}(\bar{z})=\frac{1}{8 \pi i} \oint d z \; \mathcal{P}(z, \bar{z}).
    \end{split}
\end{equation}

We now define a current corresponding to the conformal operator of the gluon $A_{~\mu}^{a}$ which would generate the $\mathrm{u}(N)$-gauge transformations as
\begin{equation}
 \begin{aligned}
&G^{a}(z) = \frac{1}{2 \pi} \int d^{2} z^{\prime} \frac{1}{\left(z-z^{\prime}\right)^{2}} \mathcal{O}^{a}_{1,-1}(z^{\prime}, \bar{z}^{\prime}) \\
&\overline{G}^{a}(\bar{z}) =\frac{1}{2 \pi} \int d^{2} z^{\prime} \frac{1}{\left(\bar{z}-\bar{z}^{\prime}\right)^{2}} \mathcal{O}^{a}_{1,+1}(z^{\prime}, \bar{z}^{\prime}).
\end{aligned}
\label{eq:GGbardef}   
\end{equation}
The OPEs of $T,\overline{T}$ and $P,\overline{P}$ with a primary operator $\mathcal{O}^{a}_{\Delta,\ell}$ with conformal weights $(h,\bar{h})$ are already calculated in \cite{Fotopoulos:2019vac}. We record the result here for later use,
\begin{equation}
\begin{split}
    &T(z) \mathcal{O}_{\Delta, \ell}(w, \bar{w})=\frac{h}{(z-w)^{2}} \mathcal{O}_{\Delta, \ell}(w, \bar{w})+\frac{1}{z-w} \partial_{w} \mathcal{O}_{\Delta, \ell}(w, \bar{w})+\text { regular.}\\
    &\overline{T}(\bar{z}) \mathcal{O}_{\Delta, \ell}(w, \bar{w})=\frac{\bar{h}}{(\bar{z}-\bar{w})^{2}} \mathcal{O}_{\Delta, \ell}(w, \bar{w})+\frac{1}{\bar{z}-\bar{w}} \partial_{\bar{w}} \mathcal{O}_{\Delta, \ell}(w, \bar{w})+\operatorname{regular.}
\end{split}
\label{eq:TandO}
\end{equation}
For $P(z)$ we have 
\begin{equation}
 \begin{aligned}
P(z) \mathcal{O}_{\Delta, \ell}(w, \bar{w}) &=\frac{(\Delta-1)(\Delta+1)}{4 \Delta} \frac{1}{z-w} \mathcal{O}_{\Delta+1, \ell}(w, \bar{w})+\text { regular } \quad(\ell=\pm 1), \\
P(z) \mathcal{O}_{\Delta, \ell}(w, \bar{w}) &=\frac{(\Delta-1)(\Delta+2)}{4(\Delta+1)} \frac{1}{z-w} \mathcal{O}_{\Delta+1, \ell}(w, \bar{w})+\text { regular } \quad(\ell=\pm 2),
\end{aligned}
\label{eq:PandO}
\end{equation}
and similar OPEs hold for $\overline{P}(\bar{z})$ with conjugated poles.
\\\\
We now calculate the OPE of $G^a$ and $\overline{G}^a$ with spin 1 primary operators $\mathcal{O}^{b}_{\Delta,\ell}$. For $N$ such primary operators, by  \eqref{eq:Ocorscatamprel} we have
\[
\left\langle G^a(z) \prod_{n=1}^{M} \mathcal{O}^{b_n}_{\Delta_{n}, \ell_{n}}\left(z_{n}, \bar{z}_{n}\right)\right\rangle=\frac{1}{2 \pi} \int d^{2} z_{0} \frac{1}{\left(z-z_0\right)^{2}}\left\langle\mathcal{O}_{1,-1}^a(z_0,\bar{z}_0)\prod_{n=1}^{M} \mathcal{O}^{b_n}_{\Delta_{n}, \ell_{n}}\left(z_{n}, \bar{z}_{n}\right)\right\rangle.
\]
Further using the soft limit as in \cite[Eq. (3.32)]{Fan:2019emx} we have 
\begin{equation}
\begin{split}
    \Bigg{\langle}\mathcal{O}_{1,-1}^a(z_0,\bar{z}_0)\prod_{n=1}^{M} &\mathcal{O}^{b_n}_{\Delta_{n}, \ell_{n}}\left(z_{n}, \bar{z}_{n}\right)\Bigg{\rangle}\\&=\sum_{i=1}^M\sum_{c}\frac{f^{ab_ic}}{\bar{z}_0-\bar{z}_i}\left\langle\mathcal{O}_{\Delta_{1} ,\ell_{1}}^{b_{1}}\left(z_{1}, \bar{z}_{1}\right) \ldots \mathcal{O}_{\Delta_{i}, \ell_{i}}^{c}\left(z_{i}, \bar{z}_{i}\right) \ldots \mathcal{O}_{\Delta_{M}, 
    \ell_{M}}^{b_{M}}\left(z_{M}, \bar{z}_{M}\right)\right\rangle
\end{split}
\label{eq:YMsoftlim}
\end{equation}
The above relation implies that 
\[
\begin{split}
    \left\langle G^a(z) \prod_{n=1}^{M} \mathcal{O}^{b_n}_{\Delta_{n}, \ell_{n}}\left(z_{n}, \bar{z}_{n}\right)\right\rangle=\sum_{i=1}^M&\sum_{c}f^{ab_ic}\frac{1}{2 \pi} \int d^{2} z_{0} \frac{1}{\left(z-z_0\right)^{2}}\frac{1}{\bar{z}_0-\bar{z}_i}\\
    &\times\left\langle\mathcal{O}_{\Delta_{1} ,\ell_{1}}^{b_{1}}\left(z_{1}, \bar{z}_{1}\right) \ldots \mathcal{O}_{\Delta_{i}, \ell_{i}}^{c}\left(z_{i}, \bar{z}_{i}\right) \ldots \mathcal{O}_{\Delta_{M}, 
    \ell_{M}}^{b_{M}}\left(z_{M}, \bar{z}_{M}\right)\right\rangle.
\end{split}
\]
To solve the integral we use (cf. \cite[Eq. (3.7)]{Fotopoulos:2019vac})
\begin{equation}\label{deltaf}
    \partial_{z_0}\left(\frac{1}{z-z_0}\right)=\frac{1}{(z-z_0)^2},\quad \partial_{z_{0}}\left(\frac{1}{\bar{z}_{0}-\bar{z}}\right)=2 \pi \delta^{(2)}\left(z_{0}-z\right).
\end{equation}
Thus the integral after integrating by parts becomes 
\begin{equation}\label{intdeltafun1}
\begin{split}
    \int d^{2} z_{0} \frac{1}{\left(z-z_0\right)^{2}}\frac{1}{\bar{z}_0-\bar{z}_i}=-\int d^{2} z_{0} \left(\frac{1}{z-z_0}\right)\partial_{z_0}\left(\frac{1}{\bar{z}_0-\bar{z}_i}\right)=-\frac{2\pi}{z-z_i}.
\end{split}
\end{equation}
Thus we get the OPE as
\begin{equation}\label{eq:opeGos}
    \begin{split}
    \Bigg\langle G^a(z) \prod_{n=1}^{M} \mathcal{O}^{b_n}_{\Delta_{n}, \ell_{n}}&\left(z_{n}, \bar{z}_{n}\right)\Bigg\rangle\\&=-\sum_{i=1}^M\sum_{c}\frac{f^{ab_ic}}{z-z_i}\left\langle\mathcal{O}_{\Delta_{1} ,\ell_{1}}^{b_{1}}\left(z_{1}, \bar{z}_{1}\right) \ldots \mathcal{O}_{\Delta_{i}, \ell_{i}}^{c}\left(z_{i}, \bar{z}_{i}\right) \ldots \mathcal{O}_{\Delta_{M}, 
    \ell_{M}}^{b_{M}}\left(z_{M}, \bar{z}_{M}\right)\right\rangle
\end{split}
\end{equation}
This gives the OPE
\begin{equation}\label{eq:opeGO}
    G^a(z)\mathcal{O}^b_{\Delta,\ell}(w,\bar{w})=\frac{1}{w-z}\sum_{c}f^{abc}\mathcal{O}_{\Delta ,\ell}^{c}(w,\bar{w}).
\end{equation}
Similarly, we have
\begin{equation}\label{eq:opeGbaros}
    \begin{split}
    \Bigg\langle \overline{G}^a(\bar{z}) \prod_{n=1}^{M} \mathcal{O}^{b_n}_{\Delta_{n}, \ell_{n}}&\left(z_{n}, \bar{z}_{n}\right)\Bigg\rangle\\&=-\sum_{i=1}^M\sum_{c}\frac{f^{ab_ic}}{\bar{z}-\bar{z}_i}\left\langle\mathcal{O}_{\Delta_{1} ,\ell_{1}}^{b_{1}}\left(z_{1}, \bar{z}_{1}\right) \ldots \mathcal{O}_{\Delta_{i}, \ell_{i}}^{c}\left(z_{i}, \bar{z}_{i}\right) \ldots \mathcal{O}_{\Delta_{M}, 
    \ell_{M}}^{b_{M}}\left(z_{M}, \bar{z}_{M}\right)\right\rangle
\end{split}
\end{equation}
which gives the OPE
\begin{equation}\label{eq:opeGbarO}
    \overline{G}^a(\bar{z})\mathcal{O}^b_{\Delta,\ell}(w,\bar{w})=\frac{1}{\bar{w}-\bar{z}}\sum_{c}f^{abc}\mathcal{O}_{\Delta ,\ell}^{c}(w,\bar{w}).
\end{equation}
\subsection{The OPEs of symmetry current generators} \label{opeEYMBMS}
\noindent
We now have all the required tools for computing the OPEs of the symmetry currents. The OPEs of combinations of $T,\overline{T}$ and $P,\overline{P}$ have already been computed in \cite{Fotopoulos:2019vac}. We compute the remaining combinations below.
\subsubsection{$G^{a}G^{b}$ and $\overline{G}^{a}\overline{G}^{b}$}\label{sec:GG}
We have
\[
\begin{split}
     \left\langle G^a(z) G^b(w) \prod_{n=2}^{M}\mathcal{O}^{b_n}_{\Delta_{n}, \ell_{n}}\left(z_{n}, \bar{z}_{n}\right)\right\rangle &= \frac{1}{4\pi^2}\int d^{2} z_{0} \frac{1}{\left(z-z_{0}\right)^{2}} \int d^{2} z_{1} \frac{1}{\left(w-z_{1}\right)^{2}}\\
    & \times \left\langle\mathcal{O}_{1,-1}^a(z_0,\bar{z}_0)\mathcal{O}_{1,-1}^b(z_1,\bar{z}_1)\prod_{n=1}^{M} \mathcal{O}^{b_n}_{\Delta_{n}, \ell_{n}}\left(z_{n}, \bar{z}_{n}\right)\right\rangle.
\end{split}
\]
Since this OPE involves double soft limit with identical helicity, it does not depend on the order in which take the soft limit. We first take the soft limit corresponding to the gauge index $a$. We have
\[
\begin{split}
    \Bigg{\langle}\mathcal{O}_{1,-1}^a(z_0,\bar{z}_0)\mathcal{O}_{1,-1}^b(z_1,\bar{z}_1)&\prod_{n=2}^{M} \mathcal{O}^{b_n}_{\Delta_{n}, \ell_{n}}\left(z_{n}, \bar{z}_{n}\right)\Bigg{\rangle}\\&=\sum_{i=1}^M\sum_{c}\frac{f^{ab_ic}}{\bar{z}_0-\bar{z}_i}\left\langle\mathcal{O}_{1 ,-1}^{b_{1}=b}\left(z_{1}, \bar{z}_{1}\right) \ldots \mathcal{O}_{\Delta_{i}, \ell_{i}}^{c}\left(z_{i}, \bar{z}_{i}\right) \ldots \mathcal{O}_{\Delta_{M}, 
    \ell_{M}}^{b_{M}}\left(z_{M}, \bar{z}_{M}\right)\right\rangle.
\end{split}
\]
We now take the first shadow transform. Using Eq. \eqref{intdeltafun1}, we get 
\[
\begin{split}
     \left\langle G^a(z) G^b(w) \prod_{n=2}^{M}\mathcal{O}^{b_n}_{\Delta_{n}, \ell_{n}}\left(z_{n}, \bar{z}_{n}\right)\right\rangle
    &=- \frac{1}{2\pi}\sum_{i=1}^M\sum_{c}f^{ab_ic}\int d^{2} z_{1} \frac{1}{\left(w-z_{1}\right)^{2}}\frac{1}{z-z_i}\\
    & \times \left\langle\mathcal{O}_{1 ,-1}^{b_{1}=b}\left(z_{1}, \bar{z}_{1}\right) \ldots \mathcal{O}_{\Delta_{i}, \ell_{i}}^{c}\left(z_{i}, \bar{z}_{i}\right) \ldots \mathcal{O}_{\Delta_{M}, 
    \ell_{M}}^{b_{M}}\left(z_{M}, \bar{z}_{M}\right)\right\rangle.
\end{split}
\]
The integral on the right hand side is regular as $z\to w$ as a result of the global conformal invariance of the correlator in the integrand, see Appendix \ref{appA1}.

This implies that the concerned OPE is regular. Hence,
\begin{equation}
G^{a}(z) G^{b}(w) \sim \text { regular.}
\label{eq:opeGG}
\end{equation}
Similarly,
\begin{equation}
\overline{G}^{a}(\bar{z}) \overline{G}^{b}(\bar{w}) \sim \text{regular.}
\label{eq:opeGbarGbar}
\end{equation}

\subsubsection{$G^{a}\overline{G}^{b},\overline{G}^{a}G^{b}$ and the Composite Current}
We consider
\[
\begin{split}
     \left\langle G^{a}(z) \overline{G}^{b}(\bar{w}) \prod_{n=2}^{M}\mathcal{O}^{b_n}_{\Delta_{n}, \ell_{n}}\left(z_{n}, \bar{z}_{n}\right)\right\rangle &=\frac{1}{4\pi^2}\int d^{2} z_{0} \frac{1}{\left(z-z_{0}\right)^{2}} \int d^{2} z_{1} \frac{1}{\left(\bar{w}-\bar{z}_{1}\right)^{2}}\\
    & \times \left\langle\mathcal{O}_{1,-1}^a(z_0,\bar{z}_0)\mathcal{O}_{1,+1}^b(z_1,\bar{z}_1)\prod_{n=2}^{M} \mathcal{O}^{b_n}_{\Delta_{n}, \ell_{n}}\left(z_{n}, \bar{z}_{n}\right)\right\rangle.
\end{split}
\]
This is a double soft limit with opposite helicities and hence we expect that the result to depend on the order in which we take the soft limits. Let us first take the soft limit corresponding to the gauge index $a$. Using the soft limit, 
\begin{equation}
\begin{split}\label{eq01}
    \Bigg{\langle}\mathcal{O}_{1,-1}^a(z_0,\bar{z}_0)&\mathcal{O}_{1,+1}^b(z_1,\bar{z}_1)\prod_{n=2}^{M} \mathcal{O}^{b_n}_{\Delta_{n}, \ell_{n}}\left(z_{n}, \bar{z}_{n}\right)\Bigg{\rangle}\\
    &=\sum_{c}\frac{f^{ab c}}{\bar{z}_0-\bar{z}_1}\left\langle\mathcal{O}_{1 ,+1}^{c}\left(z_{1}, \bar{z}_{1}\right) \ldots \mathcal{O}_{\Delta_{M}, 
    \ell_{M}}^{b_{M}}\left(z_{M}, \bar{z}_{M}\right)\right\rangle\\
    &+ \sum_{i=2}^M\sum_{c}\frac{f^{ab_ic}}{\bar{z}_0-\bar{z}_i}\left\langle\mathcal{O}_{1 ,+1}^{b_1=b}\left(z_{1}, \bar{z}_{1}\right) \ldots \mathcal{O}_{\Delta_{i}, \ell_{i}}^{c}\left(z_{i}, \bar{z}_{i}\right) \ldots \mathcal{O}_{\Delta_{M}, 
    \ell_{M}}^{b_{M}}\left(z_{M}, \bar{z}_{M}\right)\right\rangle.
    \\&=\sum_{i=1}^M\sum_{c}\frac{f^{ab_ic}}{\bar{z}_0-\bar{z}_i}\left\langle\mathcal{O}_{1 ,+1}^{b_{1}=b}\left(z_{1}, \bar{z}_{1}\right) \ldots \mathcal{O}_{\Delta_{i}, \ell_{i}}^{c}\left(z_{i}, \bar{z}_{i}\right) \ldots \mathcal{O}_{\Delta_{M}, 
    \ell_{M}}^{b_{M}}\left(z_{M}, \bar{z}_{M}\right)\right\rangle.
\end{split}
\end{equation}
and following the calculations of previous section, we have 
\[
\begin{split}
     \left\langle G^{a}(z) \overline{G}^{b}(\bar{w}) \prod_{n=2}^{M}\mathcal{O}^{b_n}_{\Delta_{n}, \ell_{n}}\left(z_{n}, \bar{z}_{n}\right)\right\rangle &=- \frac{1}{2\pi}\sum_{i=1}^M\sum_{c}f^{ab_ic}\int d^{2} z_{1} \frac{1}{\left(\bar{w}-\bar{z}_{1}\right)^{2}}\frac{1}{z-z_i}\\
    & \times \left\langle\mathcal{O}_{1 ,+1}^{b_{1}=b}\left(z_{1}, \bar{z}_{1}\right) \ldots \mathcal{O}_{\Delta_{i}, \ell_{i}}^{c}\left(z_{i}, \bar{z}_{i}\right) \ldots \mathcal{O}_{\Delta_{M}, 
    \ell_{M}}^{b_{M}}\left(z_{M}, \bar{z}_{M}\right)\right\rangle.
\end{split}
\]
We now take the second soft limit. We finally get
\[
\begin{split}
     \left\langle G^a(z) \overline{G}^b(\bar{w}) \prod_{n=2}^{M}\mathcal{O}_{\Delta_{n}, \ell_{n}}\left(z_{n}, \bar{z}_{n}\right)\right\rangle =- \frac{1}{2\pi}\sum_{i=1}^M\sum_{j=2}^M\sum_{c,d}f^{ab_ic}f^{b_{i1}b_{ij}d}\int d^{2} z_{1} \frac{1}{\left(\bar{w}-\bar{z}_{1}\right)^{2}}\frac{1}{z-z_i}\frac{1}{z_1-z_j}\\
     \times \left\langle\mathcal{O}_{\Delta_2 ,\ell_2}^{b_{i2}}\left(z_{2}, \bar{z}_{2}\right) \ldots \mathcal{O}_{\Delta_{i}, \ell_{i}}^{c}\left(z_{i}, \bar{z}_{i}\right) \ldots\ldots \mathcal{O}_{\Delta_{j}, \ell_{j}}^{d}\left(z_{j}, \bar{z}_{j}\right) \ldots \mathcal{O}_{\Delta_{M}, 
    \ell_{M}}^{b_{iM}}\left(z_{M}, \bar{z}_{M}\right)\right\rangle,
\end{split}
\]
where for $1\leq i\leq M ,\;  2 \leq j\leq M$, we have
\[
b_{ij}=\begin{cases}b_j&i\neq j\\c&i=j,\end{cases}\quad b_1=b.
\]
Let us briefly explain various terms in the above expressions. Splitting the sum over $i$ , the $i=1$ term comes from the first term of the rhs of \eqref{eq01} and here $b_{11}=c$. The $i\geq 2$ terms come from the second term of \eqref{eq01}. Further the integral is easily seen to be,

\begin{equation}\label{intineq1}
\begin{split}
    \int d^{2} z_{1} \frac{1}{\left(\bar{w}-\bar{z}_{1}\right)^{2}}\frac{1}{z-z_i}\frac{1}{z_1-z_j}&=-\frac{1}{z-z_i}\int d^{2} z_{1} \frac{1}{\left(\bar{w}-\bar{z}_{1}\right)}\partial_{\bar{z}_1}\left(\frac{1}{z_1-z_j}\right)\\&=-\frac{2\pi}{z-z_i}\frac{1}{\bar{w}-\bar{z}_{j}},\quad\quad (i\neq 1).
\end{split}
\end{equation}
When $i=1$, we have 
\begin{equation}\label{intieq1}
\begin{split}
    \int d^{2} z_{1} \frac{1}{\left(\bar{w}-\bar{z}_{1}\right)^{2}}\frac{1}{z-z_1}\frac{1}{z_1-z_j}&=-\frac{2\pi}{z-z_j}\int d^{2} z_{1} \frac{1}{\left(\bar{w}-\bar{z}_{1}\right)}\partial_{\bar{z}_1}\left(\frac{1}{z-z_1}+\frac{1}{z_1-z_j}\right)\\&=-\frac{2\pi}{z-z_j}\left(\frac{1}{\bar{w}-\bar{z}}+\frac{1}{\bar{w}-\bar{z}_{j}}\right)
\end{split}
\end{equation}
Substituting these integrals, we get 
\begin{equation}
\begin{split}
     \Bigg{\langle} G^a(z) &\overline{G}^b(\bar{w}) \prod_{n=2}^{M}\mathcal{O}^{b_n}_{\Delta_{n}, \ell_{n}}\left(z_{n}, \bar{z}_{n}\right)\Bigg{\rangle}\\&=\left[\frac{1}{\bar{w}-\bar{z}}\sum_{j=2}^M\sum_{c,d}f^{abc}f^{cb_jd}\frac{1}{z-z_j}+\sum_{j=2}^M\sum_{c,d}f^{abc}f^{cb_jd}\frac{1}{z-z_j}\frac{1}{\bar{w}-\bar{z}_{j}}\right]\\&\hspace{5.5cm}\times\left\langle\mathcal{O}_{\Delta_2,\ell_2}^{b_2}\left(z_{2}, \bar{z}_{2}\right) \ldots \mathcal{O}_{\Delta_{j}, \ell_{j}}^{d}\left(z_{j}, \bar{z}_{j}\right) \ldots \mathcal{O}_{\Delta_{M}, 
    \ell_{M}}^{b_{M}}\left(z_{M}, \bar{z}_{M}\right)\right\rangle\\& +\sum_{i=2}^M\sum_{j=2}^M\sum_{c,d}f^{ab_ic}f^{bb_{ij}d}\frac{1}{z-z_i}\frac{1}{\bar{w}-\bar{z}_{j}}\left\langle\mathcal{O}_{\Delta_2 ,\ell_2}^{b_{i2}}\left(z_{2}, \bar{z}_{2}\right) \ldots\right\rangle.
\end{split}
\end{equation}
Next we compute $\overline{G}^b(\bar{z})G^a(w)$. Proceeding as in the previous calculation we get 
\[
\begin{split}
     \Bigg{\langle} \overline{G}^b(\bar{z})&G^a(w) \prod_{n=2}^{M}\mathcal{O}_{\Delta_{n}, \ell_{n}}\left(z_{n}, \bar{z}_{n}\right)\Bigg{\rangle}=- \frac{1}{2\pi}\sum_{i=1}^M\sum_{c}f^{bb_ic}\int d^{2} z_{1} \frac{1}{\left(w-z_{1}\right)^{2}}\frac{1}{\bar{z}-\bar{z}_i}\\&\hspace{5.5cm} \times \left\langle\mathcal{O}_{1 ,-1}^{b_{1}=a}\left(z_{1}, \bar{z}_{1}\right) \ldots \mathcal{O}_{\Delta_{i}, \ell_{i}}^{c}\left(z_{i}, \bar{z}_{i}\right) \ldots \mathcal{O}_{\Delta_{M}, 
    \ell_{M}}^{b_{M}}\left(z_{M}, \bar{z}_{M}\right)\right\rangle\\&=\left[\frac{1}{w-z}\sum_{j=2}^M\sum_{c,d}f^{bac}f^{cb_jd}\frac{1}{\bar{z}-\bar{z}_j}+\sum_{j=2}^M\sum_{c,d}f^{bac}f^{cb_jd}\frac{1}{\bar{z}-\bar{z}_j}\frac{1}{w-z_{j}}\right]\\&\hspace{5.5cm}\times\left\langle\mathcal{O}_{\Delta_2,\ell_2}^{b_2}\left(z_{2}, \bar{z}_{2}\right) \ldots \mathcal{O}_{\Delta_{j}, \ell_{j}}^{d}\left(z_{j}, \bar{z}_{j}\right) \ldots \mathcal{O}_{\Delta_{M}, 
    \ell_{M}}^{b_{M}}\left(z_{M}, \bar{z}_{M}\right)\right\rangle\\& +\sum_{i=2}^M\sum_{j=2}^M\sum_{c,d}f^{bb_ic}f^{ab_{ij}d}\frac{1}{\bar{z}-\bar{z}_i}\frac{1}{w-z_{j}}\left\langle\mathcal{O}_{\Delta_2 ,\ell_2}^{b_{i2}}\left(z_{2}, \bar{z}_{2}\right) \ldots\right\rangle.
\end{split}
\]
As expected these OPEs depend on the choice of order in which we take the soft limits. To construct a quantity which is independent of this choice, we consider the following combination,
\[
\mathcal{G}^{ab}(z,\bar{z};w,\bar{w}):=G^{a}(z)\overline{G}^{b}(\bar{w})-\overline{G}^{b}(\bar{w}) G^{a}(z) \equiv [G^{a}(z),\overline{G}^{b}(\bar{w})].
\]
Let us now compute the OPE
\[
\begin{split}
    \Bigg<  \mathcal{G}^{ab}&(z,\bar{z};w,\bar{w}) \prod_{n=2}^{M}\mathcal{O}_{\Delta_{n}, \ell_{n}}\left(z_{n}, \bar{z}_{n}\right)\Bigg>\\
    &=\sum_{j=2}^M\sum_{c,d}f^{abc}f^{cb_jd}\left(\frac{1}{\bar{w}-\bar{z}}\frac{1}{z-z_j}+\frac{1}{z-w}\frac{1}{\bar{w}-\bar{z}_j}\right)\left\langle \ldots \mathcal{O}_{\Delta_{j}, \ell_{j}}^{d}\left(z_{j}, \bar{z}_{j}\right) \ldots \right\rangle\\&+\sum_{j=2}^M\sum_{c,d}f^{abc}f^{cb_jd}\left( \frac{1}{z-z_j}\frac{1}{\bar{w}-\bar{z}_{j}}+\frac{1}{\bar{w}-\bar{z}_j}\frac{1}{z-z_{j}}\right)\left\langle \ldots \mathcal{O}_{\Delta_{j}, \ell_{j}}^{d}\left(z_{j}, \bar{z}_{j}\right) \ldots \right\rangle\\
    &+\sum_{i=2}^M\sum_{j=2}^M\sum_{c,d}\left(f^{ab_ic}f^{bb_{ij}d}\frac{1}{z-z_i}\frac{1}{\bar{w}-\bar{z}_{j}}-f^{bb_ic}f^{ab_{ij}d}\frac{1}{\bar{w}-\bar{z}_i}\frac{1}{z-z_{j}}\right)\\&\hspace{6.2cm}\times\left\langle \ldots\mathcal{O}_{\Delta_{i}, \ell_{i}}^{c}\left(z_{i}, \bar{z}_{i}\right) \ldots \mathcal{O}_{\Delta_{j}, \ell_{j}}^{d}\left(z_{j}, \bar{z}_{j}\right) \ldots \right\rangle,
\end{split}
\]
where we used the fact that $f^{abc}=-f^{bac}.$

Recollecting terms after a bit of simplification, we have
\[
\begin{split}
    \Bigg<  \mathcal{G}^{ab}&(z,\bar{z};w,\bar{w}) \prod_{n=2}^{M}\mathcal{O}_{\Delta_{n}, \ell_{n}}\left(z_{n}, \bar{z}_{n}\right)\Bigg>\\&=\sum_{j=2}^M\sum_{c,d}f^{abc}f^{cb_jd}\left(\frac{1}{\bar{w}-\bar{z}}\frac{1}{z-z_j}+\frac{1}{z-w}\frac{1}{\bar{w}-\bar{z}_j}\right)\left\langle \ldots \mathcal{O}_{\Delta_{j}, \ell_{j}}^{d}\left(z_{j}, \bar{z}_{j}\right) \ldots \right\rangle\\&+\sum_{j=2}^M\sum_{c,d}\left[ \frac{(f^{abc}f^{cb_jd} + f^{ab_j c}f^{bc d})}{(z-z_j)(\bar{w}-\bar{z}_{j})}+\frac{(f^{abc}f^{cb_jd}- f^{bb_j c}f^{ac d})}{(\bar{w}-\bar{z}_j)(z-z_{j})}\right]\left\langle \ldots \mathcal{O}_{\Delta_{j}, \ell_{j}}^{d}\left(z_{j}, \bar{z}_{j}\right) \ldots \right\rangle\\
    &+\sum_{\substack{i,j=2\\i\neq j}}^M\sum_{c,d}\left(f^{ab_ic}f^{bb_{j}d}\frac{1}{z-z_i}\frac{1}{\bar{w}-\bar{z}_{j}}-f^{bb_ic}f^{ab_{j}d}\frac{1}{\bar{w}-\bar{z}_i}\frac{1}{z-z_{j}}\right)\left\langle\mathcal{O}_{\Delta_2 ,\ell_2}^{b_{i2}}\left(z_{2}, \bar{z}_{2}\right) \ldots\right\rangle
\end{split}
\]
The third term on the r.h.s of the above equation exactly vanishes on swapping dummy summation variables $c\leftrightarrow d$ and $i\leftrightarrow j$.

Now using  the Jacobi identity for structure constants,
$$f^{abc}f^{ced}+f^{aec}f^{bcd}+f^{bec}f^{cad}=0
$$
we get ,
$$
\sum_c(f^{abc}f^{cb_jd} + f^{ab_j c}f^{bc d})=\sum_c(f^{abc}f^{cb_jd} + f^{b_jac}f^{cb d})=-\sum_c f^{bb_jc}f^{cad}.
$$
\text{Similarly,}
\begin{equation}
\sum_c (f^{abc}f^{cb_jd}- f^{bb_j c}f^{ac d})=-\sum_c f^{b_jac}f^{cb d}.
\label{eq:stconstrel2}    
\end{equation}

Substituting these we get,
\begin{equation}\label{newG}
\begin{split}
    \Bigg<  \mathcal{G}^{ab}&(z,\bar{z};w,\bar{w}) \prod_{n=2}^{M}\mathcal{O}_{\Delta_{n}, \ell_{n}}\left(z_{n}, \bar{z}_{n}\right)\Bigg>\\&=\sum_{j=2}^M\sum_{c,d}f^{abc}f^{cb_jd}\left(\frac{1}{\bar{w}-\bar{z}}\frac{1}{z-z_j}+\frac{1}{z-w}\frac{1}{\bar{w}-\bar{z}_j}\right)\left\langle \ldots \mathcal{O}_{\Delta_{j}, \ell_{j}}^{d}\left(z_{j}, \bar{z}_{j}\right) \ldots \right\rangle\\&+\sum_{j=2}^M\sum_{c,d}\left[ \frac{f^{bb_jc}f^{acd}}{(z-z_j)(\bar{w}-\bar{z}_{j})}+\frac{f^{b_jac}f^{bc d}}{(\bar{w}-\bar{z}_j)(z-z_{j})}\right]\left\langle \ldots \mathcal{O}_{\Delta_{j}, \ell_{j}}^{d}\left(z_{j}, \bar{z}_{j}\right) \ldots \right\rangle
\end{split}    
\end{equation}

Here the singular terms in the above correlator involve insertions of the operator $\mathcal{O}_{1 ,-1}^{c}\left(z, \bar{z}\right)$ and $\mathcal{O}_{1 ,+1}^{c}\left(z, \bar{z}\right)$ with shadow transformation which are $G^{c}(z)$ and $\overline{G}^{c}(\bar{z})$.
We have the OPE corresponding to our defined current,
\begin{equation}
    \begin{split}
     \Bigg<  \mathcal{G}^{ab}&(z,\bar{z};w,\bar{w}) \prod_{n=2}^{M}\mathcal{O}_{\Delta_{n}, \ell_{n}}\left(z_{n}, \bar{z}_{n}\right)\Bigg>\\&=\sum_{j=2}^M\sum_{c,d}f^{abc}f^{cb_jd}\left(\frac{1}{\bar{w}-\bar{z}}\frac{1}{z-z_j}+\frac{1}{z-w}\frac{1}{\bar{w}-\bar{z}_j}\right)\left\langle \ldots \mathcal{O}_{\Delta_{j}, \ell_{j}}^{d}\left(z_{j}, \bar{z}_{j}\right) \ldots \right\rangle\\
     &= \sum_{c,d}\frac{f^{abc}}{\bar{z}-\bar{w}} \Bigg[-\sum_{j=2}^M  \frac{f^{cb_jd}}{z-z_j}\left\langle \ldots \mathcal{O}_{\Delta_{j}, \ell_{j}}^{d}\left(z_{j}, \bar{z}_{j}\right) \ldots \right\rangle \Bigg]\\
     &\hspace{4.2cm} -\sum_{c,d}\frac{f^{abc}}{z-w} \Bigg[-\sum_{j=2}^M  \frac{f^{cb_jd}}{\bar{w}-\bar{z}_j}\left\langle \ldots \mathcal{O}_{\Delta_{j}, \ell_{j}}^{d}\left(z_{j}, \bar{z}_{j}\right) \ldots \right\rangle\Bigg]
    \end{split}
\end{equation}
Using Eq.(\ref{eq:opeGos}) and Eq.(\ref{eq:opeGbaros}), we can write the bracketed terms in r.h.s as
\[
\begin{split}
    &-\sum_{j=2}^M  \frac{f^{cb_jd}}{z-z_j}\left\langle \ldots \mathcal{O}_{\Delta_{j}, \ell_{j}}^{d}\left(z_{j}, \bar{z}_{j}\right) \ldots \right\rangle=\Bigg\langle G^c(z) \prod_{n=2}^{M} \mathcal{O}^{b_n}_{\Delta_{n}, \ell_{n}}\left(z_{n}, \bar{z}_{n}\right)\Bigg\rangle,  \\
 &-\sum_{j=2}^M  \frac{f^{cb_jd}}{\bar{w}-\bar{z}_j}\left\langle \ldots \mathcal{O}_{\Delta_{j}, \ell_{j}}^{d}\left(z_{j}, \bar{z}_{j}\right) \ldots \right\rangle=\Bigg\langle \overline{G}^c(\bar{w}) \prod_{n=2}^{M} \mathcal{O}^{b_n}_{\Delta_{n}, \ell_{n}}\left(z_{n}, \bar{z}_{n}\right)\Bigg\rangle.
\end{split}
\]
Thus,
\begin{equation}
    \begin{split}
        \Bigg<  \mathcal{G}^{ab}(z,\bar{z};w,\bar{w}) \prod_{n=2}^{M}\mathcal{O}_{\Delta_{n}, \ell_{n}}\left(z_{n}, \bar{z}_{n}\right)\Bigg>&= \sum_{c,d}\frac{f^{abc}}{\bar{z}-\bar{w}} \Bigg\langle G^c(z) \prod_{n=2}^{M} \mathcal{O}^{b_n}_{\Delta_{n}, \ell_{n}}\left(z_{n}, \bar{z}_{n}\right)\Bigg\rangle \\
        &\hspace{1.5cm}-\sum_{c,d}\frac{f^{abc}}{z-w} \Bigg\langle \overline{G}^c(\bar{w}) \prod_{n=2}^{M} \mathcal{O}^{b_n}_{\Delta_{n}, \ell_{n}}\left(z_{n}, \bar{z}_{n}\right)\Bigg\rangle
    \end{split}
\end{equation}
Hence the OPE:
\[
\begin{split}
    G^{a}(z)\overline{G}^{b}(\bar{w})&-\overline{G}^{b}(\bar{w}) G^{a}(z) = \; \frac{1}{\bar{z}-\bar{w}}\sum_cf^{abc}G^{c}(z) +\frac{1}{w-z}\sum_cf^{abc}\overline{G}^{c}(\bar{w}) + \text{regular.}
\end{split}
\]
We further extract the finite piece of $\mathcal{G}^{ab}(z,\bar{z};w,\bar{w})$ remaining at $w=z$. To this end, let us define the current 
\begin{equation}
    \mathcal{G}^{ab}(z,\bar{z})=~:\mathcal{G}^{ab}(z,\bar{z};z,\bar{z}):~=~:G^{a}(z)\overline{G}^{b}(\bar{z})-\overline{G}^{b}(\bar{z}) G^{a}(z):~\equiv ~:[G^{a}(z),\overline{G}^{b}(\bar{z})]:.
\end{equation}
Hence the correlator of the normal ordered current is,
\begin{equation}\label{eq:opecurlyGos}
\begin{split}
    \Bigg{\langle} \mathcal{G}^{ab}(z,\bar{z})\prod_{n=2}^{M}\mathcal{O}^{b_n}_{\Delta_{n}, \ell_{n}}\left(z_{n},\bar{z}_{n}\right)\Bigg{\rangle}&=\sum_{j=2}^M\sum_{c,d}\left[\frac{f^{bb_jc}f^{acd}+f^{b_jac}f^{bc d}}{(z-z_j)(\bar{z}-\bar{z}_{j})}\right]\left\langle \ldots \mathcal{O}_{\Delta_{j}, \ell_{j}}^{d}\left(z_{j}, \bar{z}_{j}\right) \ldots \right\rangle\\
    &=\sum_{j=2}^M\sum_{c,d}\left[\frac{f^{abc}f^{cb_jd}}{(z-z_j)(\bar{z}-\bar{z}_{j})}\right]\left\langle \ldots \mathcal{O}_{\Delta_{j}, \ell_{j}}^{d}\left(z_{j}, \bar{z}_{j}\right) \ldots \right\rangle,
\end{split}
\end{equation}
where we used Eq. \eqref{eq:stconstrel2}. In particular, we have the OPE 
\begin{equation}\label{eq:opecurlyGO}
    \mathcal{G}^{ab}(z,\bar{z})\mathcal{O}^c_{\Delta,\ell}(w,\bar{w})=\frac{1}{z-w}\frac{1}{\bar{z}-\bar{w}}\sum_{d,e}f^{abd}f^{dce}\mathcal{O}^e_{\Delta,\ell}(w,\bar{w})+\text{regular.}
\end{equation}
\subsubsection{$TG^{a},T\overline{G}^{a}$ and $\overline{T}G^{a},\overline{T}\; \overline{G}^a$}\label{sec:TG}
\begin{itemize}
\item $T(z)G^{a}(w)$:\\

We have
\[
\begin{split}
    \bigg{\langle}T(z) G^{a}(w) &\prod_{n=2}^{M} \mathcal{O}^{b_n}_{\Delta_{n}, \ell_n}\left(z_{n}, \bar{z}_{n}\right)\bigg{\rangle}
\\&=\lim _{\Delta \rightarrow 1} \frac{1}{2 \pi} \int d^{2} z_{1} \frac{1}{\left(w-z_{1}\right)^{2}}\left\langle T(z) \mathcal{O}^{a}_{\Delta,-1}\left(z_{1}, \bar{z}_{1}\right) \prod_{n=2}^{M} \mathcal{O}^{b_n}_{\Delta_{n}, \ell_{n}}\left(z_{n}, \bar{z}_{n}\right)\right\rangle\\
&=\lim _{\Delta \rightarrow 1} \frac{1}{2 \pi} \int d^{2} z_{1} \frac{1}{\left(w-z_{1}\right)^{2}}\left[\frac{h}{\left(z-z_{1}\right)^{2}}\left\langle\mathcal{O}^{a}_{\Delta,-1}\left(z_{1}, \bar{z}_{1}\right) \prod_{n=2}^{M} \mathcal{O}^{b_n}_{\Delta_n, \ell_n}\left(z_{n}, \bar{z}_{n}\right)\right\rangle\right.\\
&\hspace{2.5cm}+\left.\frac{1}{z-z_{1}} \partial_{z_{1}}\left\langle \mathcal{O}^{a}_{\Delta,-1}\left(z_1, \bar{z}_{1}\right) \prod_{n=2}^{M} \mathcal{O}^{b_n}_{\Delta_{n}, \ell_n }\left(z_{n}, \bar{z}_{n}\right)\right\rangle\right]+\text{reg.},
\end{split}
\]
where we used \cite[Eq. (3.19)]{ Fotopoulos:2019tpe}. Now since $h=0$, the first term in the square bracket vanishes. To simplify notations, put $W= w-z_1$ and $Z=z-z_{1}$. We get 
\[
\begin{split}
    \bigg{\langle}T(z) &G^{a}(w) \prod_{n=2}^{M} \mathcal{O}^{b_n}_{\Delta_{n}, \ell_n}\left(z_{n}, \bar{z}_{n}\right)\bigg{\rangle}
\\&=\lim_{\Delta \rightarrow 1} \frac{1}{2 \pi} \int d^{2} z_{1}\frac{1}{W^2Z}\partial_{z_{1}}\left\langle \mathcal{O}^{a}_{\Delta,-1}\left(z_1, \bar{z}_{1}\right) \prod_{n=2}^{M} \mathcal{O}^{b_n}_{\Delta_{n}, \ell_n }\left(z_{n}, \bar{z}_{n}\right)\right\rangle+\text{regular}\\&=\lim_{\Delta \rightarrow 1} \frac{1}{2 \pi} \int d^{2} z_{1}\frac{1}{z-w}\left(\frac{1}{W^2}-\frac{1}{WZ}\right)\partial_{z_{1}}\left\langle \mathcal{O}^{a}_{\Delta,-1}\left(z_1, \bar{z}_{1}\right) \prod_{n=2}^{M} \mathcal{O}^{b_n}_{\Delta_{n}, \ell_n }\left(z_{n}, \bar{z}_{n}\right)\right\rangle+\text{reg.}
\end{split}
\]
Using integration by parts in the first term in the above integral, we get 
\[
\begin{split}
    \lim_{\Delta \rightarrow 1} \frac{1}{2 \pi} \int d^{2} z_{1}&\frac{1}{W^2}\partial_{z_{1}}\left\langle \mathcal{O}^{a}_{\Delta,-1}\left(z_1, \bar{z}_{1}\right) \prod_{n=2}^{M} \mathcal{O}^{b_n}_{\Delta_{n}, \ell_n }\left(z_{n}, \bar{z}_{n}\right)\right\rangle\\&=-\lim_{\Delta \rightarrow 1} \frac{1}{2 \pi} \int d^{2} z_{1}\partial_{z_1}\left(\frac{1}{W^2}\right)\left\langle \mathcal{O}^{a}_{\Delta,-1}\left(z_1, \bar{z}_{1}\right) \prod_{n=2}^{M} \mathcal{O}^{b_n}_{\Delta_{n}, \ell_n }\left(z_{n}, \bar{z}_{n}\right)\right\rangle\\&=\partial_{w}\left[\lim_{\Delta \rightarrow 1} \frac{1}{2 \pi} \int d^{2} z_{1}\frac{1}{W^2}\left\langle \mathcal{O}^{a}_{\Delta,-1}\left(z_1, \bar{z}_{1}\right) \prod_{n=2}^{M} \mathcal{O}^{b_n}_{\Delta_{n}, \ell_n }\left(z_{n}, \bar{z}_{n}\right)\right\rangle\right]\\&=\left\langle \partial_wG^a(w)\prod_{n=2}^{M} \mathcal{O}^{b_n}_{\Delta_{n}, \ell_n }\left(z_{n}, \bar{z}_{n}\right)\right\rangle,
\end{split}
\]
where we used the fact that 
\[
\partial_{z_1}\frac{1}{W^2}=-\partial_w\frac{1}{W^2}.
\]
Next, using integration by parts, we have
\[
\begin{split}
   \lim_{\Delta \rightarrow 1} \frac{1}{2 \pi} \int d^{2} z_{1}&\frac{1}{WZ}\partial_{z_{1}}\left\langle \mathcal{O}^{a}_{\Delta,-1}\left(z_1, \bar{z}_{1}\right) \prod_{n=2}^{M} \mathcal{O}^{b_n}_{\Delta_{n}, \ell_n }\left(z_{n}, \bar{z}_{n}\right)\right\rangle\\&=-\lim_{\Delta \rightarrow 1} \frac{1}{2 \pi} \int d^{2} z_{1}\left(\frac{1}{W^2Z}+\frac{1}{WZ^2}\right)\left\langle \mathcal{O}^{a}_{\Delta,-1}\left(z_1, \bar{z}_{1}\right) \prod_{n=2}^{M} \mathcal{O}^{b_n}_{\Delta_{n}, \ell_n }\left(z_{n}, \bar{z}_{n}\right)\right\rangle. \\&=\lim_{\Delta \rightarrow 1} \frac{1}{2 \pi} \int d^{2} z_{1}\left(\frac{1}{z-w}\frac{1}{WZ}-\frac{1}{WZ^2}\right)\left\langle \mathcal{O}^{a}_{\Delta,-1}\left(z_1, \bar{z}_{1}\right) \prod_{n=2}^{M} \mathcal{O}^{b_n}_{\Delta_{n}, \ell_n }\left(z_{n}, \bar{z}_{n}\right)\right\rangle\\&\hspace{6cm}-\frac{1}{z-w}\left\langle G^a(w)\prod_{n=2}^{M} \mathcal{O}^{b_n}_{\Delta_{n}, \ell_n }\left(z_{n}, \bar{z}_{n}\right)\right\rangle.
\end{split}
\]
where we have used,\[
\frac{1}{W^2Z}=\frac{1}{z-w}\left(\frac{1}{W^2}-\frac{1}{WZ}\right),
\]
\noindent
The first integral is shown to vanish in Appendix \ref{appA2} as a result of the global conformal invariance of the correlator in the integrand. Putting together everything we get
\begin{equation}
    \begin{split}
    \bigg{\langle}&T(z) G^{a}(w) \prod_{n=2}^{M} \mathcal{O}^{b_n}_{\Delta_{n}, \ell_n}\left(z_{n}, \bar{z}_{n}\right)\bigg{\rangle}\\&=\frac{1}{(z-w)^2}\bigg{\langle} G^{a}(w) \prod_{n=2}^{M} \mathcal{O}^{b_n}_{\Delta_{n}, \ell_n}\left(z_{n}, \bar{z}_{n}\right)\bigg{\rangle}+\frac{1}{z-w}\left\langle \partial_wG^a(w)\prod_{n=2}^{M} \mathcal{O}^{b_n}_{\Delta_{n}, \ell_n }\left(z_{n}, \bar{z}_{n}\right)\right\rangle.
    \end{split}
\end{equation}
This immediately gives the OPE
\begin{equation}\label{eq:opeTG}
    T(z) G^{a}(w) = \frac{1}{(z-w)^2}G^a(w)+\frac{1}{z-w} \partial_{w} G^{a}(w)+ \text{regular.}
\end{equation}
\item $\overline{T}(\bar{z})G^{a}(w)$ and $T(z)\overline{G}^{a}(\bar{w})$:

We have,
\[
\begin{split}
\bigg{\langle}\overline{T}(\bar{z}) G^{a}(w) &\prod_{n=2}^{M} \mathcal{O}^{b_n}_{\Delta_{n}, \ell_n}\left(z_{n}, \bar{z}_{n}\right)\bigg{\rangle}
\\&=\lim _{\Delta \rightarrow 1} \frac{1}{2 \pi} \int d^{2} z_{1} \frac{1}{\left(w-z_{1}\right)^{2}}\left\langle\overline{T}(\bar{z}) \mathcal{O}^{a}_{\Delta,-1}\left(z_{1}, \bar{z}_{1}\right) \prod_{n=2}^{M} \mathcal{O}^{b_n}_{\Delta_{n}, \ell_{n}}\left(z_{n}, \bar{z}_{n}\right)\right\rangle\\
&=\lim _{\Delta \rightarrow 1} \frac{1}{2 \pi} \int d^{2} z_{1} \frac{1}{\left(w-z_{1}\right)^{2}}\left[\frac{\bar{h}}{\left(\bar{z}-\bar{z}_{1}\right)^{2}}\left\langle\mathcal{O}^{a}_{\Delta,-1}\left(z_{1}, \bar{z}_{1}\right) \prod_{n=2}^{M} \mathcal{O}_{\Delta_n, \ell_n}\left(z_{n}, \bar{z}_{n}\right)\right\rangle\right.\\
&\hspace{2.5cm}+\left.\frac{1}{\bar{z}-\bar{z}_{1}} \partial_{\bar{z}_{1}}\left\langle \mathcal{O}^{a}_{\Delta,-1}\left(z_1, \bar{z}_{1}\right) \prod_{n=2}^{M} \mathcal{O}^{b_n}_{\Delta_{n}, \ell_n }\left(z_{n}, \bar{z}_{n}\right)\right\rangle\right]+\text{reg.},
\end{split}
\]
where we used \cite[Eq. (3.19)]{Fotopoulos:2019tpe}. Using integration by parts in second term and simplifying the notation by putting $W= w-z_1$ and $\bar{Z} = \bar{z}-\bar{z}_1$  we get
\[
\begin{split}
\bigg{\langle}\overline{T}&(\bar{z}) G^{a}(w) \prod_{n=2}^{M} \mathcal{O}_{\Delta_{n}, \ell_n}\left(z_{n}, \bar{z}_{n}\right)\bigg{\rangle}\\&=\lim _{\Delta \rightarrow 1} \frac{1}{2 \pi} \int d^{2} z_{1}\left[\frac{(\bar{h}-1)}{W^2 \bar{Z}^2}\left\langle \mathcal{O}^{a}_{\Delta,-1}\left(z_{1}, \bar{z}_{1}\right) \prod_{n=2}^{M} \mathcal{O}^{b_n} _{\Delta_{n}, \ell_n} \left(z_{n}, \bar{z}_{n}\right)\right\rangle\right. \\
&\hspace{4cm}-\left.\frac{1}{\bar{Z}} \left(\partial_{\bar{z}_{1}} \frac{1}{W^{2}}\right)\left\langle \mathcal{O}^{a}_{\Delta, -1}\left(z_{1}, \bar{z}_{1}\right) \prod_{n=2}^{M} \mathcal{O}^{b_n}_{\Delta_{n}, \ell_n }\left(z_{n}, \bar{z}_{n}\right)\right\rangle \right]+\text {reg.}\\
& =-\lim _{\Delta \rightarrow 1} \frac{1}{2 \pi} \int d^{2} z_{1}\; \frac{1}{\bar{Z}} \left(\partial_{\bar{z}_{1}}\frac{1}{W^{2}}\right)\left\langle \mathcal{O}^{a}_{\Delta,-1}\left(z_{1}, \bar{z}_{1}\right) \prod_{n=2}^{M} \mathcal{O}^{b_n}_{\Delta_{n}, \ell_n }\left(z_{n}, \bar{z}_{n}\right)\right\rangle+ \text{reg.}
\end{split}
\]
where we used integration by parts and the fact that $\bar{h}\to 1$ when $\Delta\to 1$.\\
Now,
\[
\begin{split}
    &\partial_{\bar{z}_{1}}\frac{1}{W^{2}}=\partial_{\bar{z}} \partial_{z_{1}}\frac{1}{W}=\partial_{z_{1}} \partial_{\bar{z}_{1}}\frac{1}{W}=-2 \pi \partial_{z_{1}} \delta^{(2)}\left(\bar{z}_{1}-\bar{w}\right)
\end{split}
\]
So we get,
\[
\begin{split}
    \langle\overline{T}(\bar{z}) G^{a}(w)\rangle
    &=\lim _{\Delta \rightarrow 1} \bar{h} \int d^{2} z_{1} \frac{1}{\bar{Z}}\; \partial_{z_{1}} \delta^{(2)}\left(\bar{z}_{1}-\bar{w}\right)\left\langle \mathcal{O}^{a}_{\Delta, -1}\left(z_{1} ,\bar{z}_{1}\right) \prod_{n=2}^{M} \mathcal{O}^{b_n}_{\Delta_n, \ell_n }\left(z_{n}, \bar{z}_{n}\right)\right\rangle\\
    &= -\lim _{\Delta \rightarrow 1} \bar{h} \int d^{2} z_{1} \frac{1}{\bar{Z}}\;\delta^{(2)}\left(\bar{z}_{1}-\bar{w}\right)\partial_{z_{1}}\left\langle \mathcal{O}^{a}_{\Delta, -1}\left(z_{1} ,\bar{z}_{1}\right) \prod_{n=2}^{M} \mathcal{O}^{b_n}_{\Delta_n, \ell_n }\left(z_{n}, \bar{z}_{n}\right)\right\rangle\\
    &=-\lim _{\Delta \rightarrow 1} \partial _{w}\left(\frac{\left\langle \mathcal{O}^{a}_{\Delta, -1}(w, \bar{w}) \prod\limits_{n=2}^{M} \mathcal{O}^{b_n}_{\Delta_n, \ell_n} \left(z_{n}, \bar{z}_{n}\right)\right\rangle}{\bar{z}-\bar{w}}\right),
\end{split}
\]
Now the soft limit in the numerator is given by:
\begin{equation}
\begin{split}
    \lim _{\Delta \rightarrow 1}\Bigg{\langle}\mathcal{O}_{\Delta,-1}^{a}(w,\bar{w})&\prod_{n=1}^{M} \mathcal{O}^{b_n}_{\Delta_{n}, \ell_{n}}\left(z_{n}, \bar{z}_{n}\right)\Bigg{\rangle}\\&=\sum_{i=1}^M\sum_{c}\frac{f^{ab_ic}}{\bar{w}-\bar{z}_i}\left\langle\mathcal{O}_{\Delta_{1} ,\ell_{1}}^{b_{1}}\left(z_{1}, \bar{z}_{1}\right) \ldots \mathcal{O}_{\Delta_{i}, \ell_{i}}^{c}\left(z_{i}, \bar{z}_{i}\right) \ldots \mathcal{O}_{\Delta_{n}, 
    \ell_{n}}^{b_{n}}\left(z_{n}, \bar{z}_{n}\right)\right\rangle
\end{split}
\label{eq:YMsoftlim}
\end{equation}
Thus we get, \[
\begin{split}
\left\langle\overline{T}(\bar{z}) G^{a}(w) \prod_{n=2}^{M} \mathcal{O}^{b_n}_{\Delta_{n}, \ell_n }\left(z_{n}, \bar{z}_{n}\right)\right\rangle
&=-\partial_{w}\Bigg[ \sum_{i=2}^M\sum_{c}\frac{f^{ab_ic}}{(\bar{w}-\bar{z}_i)(\bar{z}-\bar{w})} \Bigg]\\
&\times
\left\langle\mathcal{O}_{\Delta_{1} ,\ell_{2}}^{b_{2}}\left(z_{1}, \bar{z}_{1}\right) \ldots \mathcal{O}_{\Delta_{i}, \ell_{i}}^{c}\left(z_{i}, \bar{z}_{i}\right) \ldots \mathcal{O}_{\Delta_{n}, 
    \ell_{n}}^{b_{n}}\left(z_{n}, \bar{z}_{n}\right)\right\rangle.
\end{split}
\]
It is now clear that we will get delta functions $\delta^{(2)}(z-w), \; \delta^{(2)}(z_i-w), $ and the OPE will be localised at $z=w,\; w=z_i$ respectively and each term will be regular as $z\to w$.\\
Thus we conclude that,
\begin{equation}
    \overline{T}(\bar{z}) G^{a}(w) \sim \text { regular.}
    \label{eq:TbarG}
\end{equation}
Similarly, 
\begin{equation}
    T(z) \overline{G}^{a}(\bar{w}) \sim \text { regular.}
    \label{eq:TGbar}
\end{equation}
\item $\overline{T}(\bar{z}) \overline{G}^{a}(\bar{w})$:
We have,
\[
\begin{split}
\bigg{\langle}\overline{T}(\bar{z}) \overline{G}^{a}(\bar{w}) &\prod_{n=2}^{M} \mathcal{O}^{b_n}_{\Delta_{n}, \ell_n}\left(z_{n}, \bar{z}_{n}\right)\bigg{\rangle}
\\&=\lim _{\Delta \rightarrow 1} \frac{1}{2 \pi} \int d^{2} z_{1} \frac{1}{\left(\bar{w}-\bar{z}_{1}\right)^{2}}\left\langle\overline{T}(\bar{z}) \mathcal{O}^{a}_{\Delta,+1}\left(z_{1}, \bar{z}_{1}\right) \prod_{n=2}^{M} \mathcal{O}^{b_n}_{\Delta_{n}, \ell_{n}}\left(z_{n}, \bar{z}_{n}\right)\right\rangle\\
&=\lim _{\Delta \rightarrow 1} \frac{1}{2 \pi} \int d^{2} z_{1} \frac{1}{\left(\bar{w}-\bar{z}_{1}\right)^{2}}\left[\frac{\bar{h}}{\left(\bar{z}-\bar{z}_{1}\right)^{2}}\left\langle\mathcal{O}^{a}_{\Delta,+1}\left(z_{1}, \bar{z}_{1}\right) \prod_{n=2}^{M} \mathcal{O}_{\Delta_n, \ell_n}\left(z_{n}, \bar{z}_{n}\right)\right\rangle\right.\\
&\hspace{2.5cm}+\left.\frac{1}{\bar{z}-\bar{z}_{1}} \partial_{\bar{z}_{1}}\left\langle \mathcal{O}^{a}_{\Delta,+1}\left(z_1, \bar{z}_{1}\right) \prod_{n=2}^{M} \mathcal{O}^{b_n}_{\Delta_{n}, \ell_n }\left(z_{n}, \bar{z}_{n}\right)\right\rangle\right]+\text{reg.},
\end{split}
\]
\noindent
Since $\Delta \to 1$ implies $\bar{h} \to 0$, following previous calculations, we see that
\begin{equation}
    \begin{split}
    \bigg{\langle}\overline{T}(\bar{z}) \overline{G}^{a}(\bar{w})\prod_{n=2}^{M} \mathcal{O}^{b_n}_{\Delta_{n}, \ell_n}\left(z_{n}, \bar{z}_{n}\right)\bigg{\rangle}&=\frac{1}{(\bar{z}-\bar{w})^2}\bigg{\langle} \overline{G}^{a}(\bar{w}) \prod_{n=2}^{M} \mathcal{O}^{b_n}_{\Delta_{n}, \ell_n}\left(z_{n}, \bar{z}_{n}\right)\bigg{\rangle}\\
    &\hspace{6ex}+\frac{1}{\bar{z}-\bar{w}}\left\langle \partial_{\bar{w}}\overline{G}^{a}(\bar{w})\prod_{n=2}^{M} \mathcal{O}^{b_n}_{\Delta_{n}, \ell_n }\left(z_{n}, \bar{z}_{n}\right)\right\rangle.
    \end{split}
\end{equation}
This implies the OPE as
\begin{equation}\label{eq:opeTbarGbar}
    \overline{T}(\bar{z}) \overline{G}^{a}(\bar{w}) =\frac{1}{(\bar{z}-\bar{w})^2}\overline{G}^a(\bar{w})+ \frac{1}{\bar{z}-\bar{w}} \partial_{\bar{w}} \overline{G}^{a}(\bar{w})+\text{regular.}
\end{equation}
\end{itemize}
\subsubsection{$PG^{a},P\overline{G}^{a}$}\label{sec:PG}
We have,
\[
\begin{split}
    P(z) G^{a}(w)=\lim _{\Delta \rightarrow 1} \frac{1}{2 \pi} \int d^{2} z^{\prime} \frac{1}{\left(w-z^{\prime}\right)^{2}} P(z) \mathcal{O}^{a}_{\Delta,-1}\left(z^{\prime}, \bar{z}^{\prime}\right)
\end{split}
\]
Now by \cite{Fotopoulos:2019tpe}, $\mathcal{O}^{a}_{\Delta, -1}$ is a primary field with conformal weight $h= \frac{\Delta -1}{2}, \; \bar{h}=\frac{\Delta+1}{2}$.
Thus using \cite[Eq. (3.17)]{Fotopoulos:2019vac}, we have
\[
\begin{split}
    P(z) G^{a}(w)&=\lim _{\Delta \rightarrow 1} \frac{1}{2 \pi} \int d^{2} z^{\prime} \frac{1}{\left(w-z^{\prime}\right)^{2}}\left[\frac{(\Delta-1)(\Delta+1)}{4 \Delta} \frac{1}{(z-z^{\prime})} \mathcal{O}^{a}_{\Delta+1,-1}\left(z^{\prime}, \bar{z}^{\prime}\right)+\text {reg.}\right]
\end{split}
\]
Hence,
\begin{equation}
    P(z) G^{a}(w) \sim \text { regular. } \label{pg}
\end{equation}
Similarly, 
\begin{equation}
    P(z) \overline{G}^{a}(\bar{w}) \sim \text { regular. } \label{pbarg}
\end{equation}
Similarly, we have 
\begin{equation}
\overline{P}(\bar{z}) \overline{G}^{a}(\bar{w}) \sim \text { regular. } \qquad \overline{P}(\bar{z}) \overline{G}^{a}(\bar{w}) \sim \text { regular. } 
    \label{pbargpbargbar}
\end{equation}
Thus we have obtained all the required OPEs for the current operators on $\mathcal{CS}^2$.
\section{Asymptotic Symmetry of EYM Theory} \label{bmsEYM}
\noindent In this section we find the asymptotic symmetry algebra of our theory. As noted earlier, we expect the algebra to be an extension of the usual $\mathfrak{bms}_4$ with $\mathrm{u}(N)$ current. Thus apart from the usual BMS algebra \cite{Fotopoulos:2019vac} generators $L_{n}, \bar{L}_{m}, P_{n-\frac{1}{2}, m-\frac{1}{2}},$ we also have $G^{a}_m, \overline{G}^{b}_n$, which are the modes of the $\mathrm{u}(N)$ currents. 
The BMS supertranslation operator $\mathcal{P}(z, \bar{z})$ is a primary operator of dimension $(\frac{3}{2}, \frac{3}{2})$ which is mode expanded as,
\begin{equation}
    \begin{split}
        \mathcal{P}(z, \bar{z}) \equiv \sum_{n, m \in \mathbb{Z}} P_{n-\frac{1}{2}, m-\frac{1}{2}} z^{-n-1} \bar{z}^{-m-1}
    \end{split}
\end{equation}
where $P_{n-\frac{1}{2}, m-\frac{1}{2}}$ are the supertranslation generators.

Since, $h=1$ for $G(z)$ and $\bar{h}=1$ for $\overline{G}(\bar{z})$, we have the following mode expansion:
\begin{equation}
    G^{a}(z)=\sum_{m \in \mathbb{Z}} \frac{G^{a}_{m}}{z^{m+1}},  \quad \overline{G}^{b}(\bar{z})=\sum_{n \in \mathbb{Z}} \frac{\overline{G}^{b}_{n}}{\bar{z}^{n+1}}
\end{equation}
where,
\begin{equation}
    \begin{split}
    G^{a}_{m}=\frac{1}{2 \pi i} \oint d z \; z^{m} G^{a}(z), \quad \overline{G}^{b}_{n}=\frac{1}{2 \pi i} \oint d \bar{z}\;  \bar{z}^{n} \overline{G}^{b}(\bar{z}) .
\end{split}
\label{eq:modesint}
\end{equation}
To this end, we use the OPEs that we have computed in the last section to get the algebra of $G_m, \overline{G}_n$.
 To find the extended algebra we compute the commutators of $G_m, \overline{G}_n$ with usual BMS algebra generators. OPEs of Eq. \eqref{pg}, \eqref{pbarg} and \eqref{pbargpbargbar} immediately implies:
\begin{equation}
    \begin{split}
    \left[G^{a}_m, P_{n-\frac{1}{2},-\frac{1}{2}}\right]=\left[\overline{G}^{a}_{m}, P_{n-\frac{1}{2},-\frac{1}{2}}\right]=\left[G^{a}_{m}, P_{-\frac{1}{2}, n-\frac{1}{2}}\right]=\left[\overline{G}^{a}_{m}, P_{-\frac{1}{2}, n-\frac{1}{2}}\right] =0.
\end{split}
\label{eq:comGPn0.5}
\end{equation} 
The remaining supertranslation generators are given by \cite{Fotopoulos:2019vac},
\begin{equation}
\begin{split}
    P_{n-\frac{1}{2}, m-\frac{1}{2}}=\frac{1}{i \pi(m+1)} \oint d \bar{w} \bar{w}^{m+1}\left[\bar{T}(\bar{w}), P_{n-\frac{1}{2},-\frac{1}{2}}\right]
\end{split}
\end{equation}
Then,
\begin{equation}
    \begin{split}
        \left[G^{a}_{m}, P_{n-\frac{1}{2}, m-\frac{1}{2}}\right]&=\frac{1}{i \pi(m+1)} \oint d \bar{w} \bar{w}^{m+1}\left[ G^{a}_{m},\left[\overline{T}(\bar{w}), P_{n-\frac{1}{2},-\frac{1}{2}}\right] \right] \\
        &=-\frac{1}{i \pi(m+1)} \oint d \bar{w} \bar{w}^{m+1}\Bigg(\left[  P_{n-\frac{1}{2},-\frac{1}{2}}, \left[ G^{a}_{m},\overline{T}(\bar{w})\right]\right] +  \left[ \overline{T}(\bar{w}),\left[P_{n-\frac{1}{2},-\frac{1}{2}}, G^{a}_{m}\right]\right]\Bigg)\\
        &=0
    \end{split}
\end{equation}
Here we have used the OPE relations of $G^a$ wtih $\bar{T}$ and the commutation relation in Eq.(\ref{eq:comGPn0.5}).\\
Similarly for the antiholomorphic current,  $ \left[\overline{G}^{a}_{m}, P_{n-\frac{1}{2}, m-\frac{1}{2}}\right] = 0$.\\

Hence we can write, 
\begin{equation}
    \begin{split}
    \left[G^{a}_{m}, P_{k, l}\right]=\left[\overline{G}^{a}_{m}, P_{k, l}\right]=0
\end{split}
\label{eq:comGnPnm}
\end{equation}
where $m, n \in \mathbb{Z} \text { and }  k, l \in \mathbb{Z}+\frac{1}{2}$.\\

The OPE of $G^{a},~\overline{G}^{a}$ with $\overline{T},~T$ respectively in Eq. \eqref{eq:TbarG}, Eq. \eqref{eq:TGbar} implies 
\begin{equation}
    \begin{split}
    \left[L_{m}, \overline{G}^{a}_{n}\right]=\left[\overline{L}_{m}, G^{a}_{n}\right]=0.
\end{split}
\label{eq:comLgbarLbarG}
\end{equation}
OPE in Eq. \eqref{eq:opeGG} and Eq. \eqref{eq:opeGbarGbar} gives 
\begin{equation}
    \left[G^{a}_m,G^{b}_n\right]=\left[\overline{G}^{a}_m,\overline{G}^{a}_n\right]=0.
    \label{G1}
\end{equation}
Next, we compute the commutator of $G^{a}_n$ with Virasoro generators. We have 
\[\begin{aligned}\left[L_{n}, G^{a}_{m}\right] &=\frac{1}{(2 \pi i)^{2}} \oint d z \oint d z^{\prime} z^{n+1} z^{\prime m}\left[T(z), G^{a}\left(z^{\prime}\right)\right] \\ &=\frac{1}{(2 \pi i)^{2}} \oint d z \oint d z^{\prime} z^{n+1} z^{\prime m} \left[\frac{1}{(z-z^{\prime})^2}  G^{a}\left(z^{\prime}\right)+\frac{1}{z-z^{\prime}} \partial_{z^{\prime}} G^{a}\left(z^{\prime}\right)\right]\\
&=\frac{1}{2 \pi i} \oint d z^{\prime} z^{\prime m} (n+1) z^{\prime n}   G^{a}\left(z^{\prime}\right)-\frac{1}{2 \pi i} \oint d z^{\prime}(m+n+1) z^{\prime m+n} G^{a}\left(z^{\prime}\right)\\
&=(n+1)G^a_{m+n}-(m+n+1)G^a_{m+n} \\ &=-m G^{a}_{m+n} \end{aligned}\]
where we used integration by parts. Thus we have
\begin{equation}
    \left[G^{a}_m, L_{n}\right]=m G^{a}_{m+n}.
\label{eq:comGL}
\end{equation}
Similarly,
\begin{equation}
    \left[\overline{G}^{a}_{m}, \overline{L}_{n}\right]=m \overline{G}^{a}_{m+n}.
    \label{eq:comGbarLbar}
\end{equation}
We also have a composite current $\mathcal{G}^{ab}(z,\bar{z})$ in our theory. 
We have the following Laurent expansion for $\mathcal{G}^{ab}(z,\bar{z})$ (see Appendix \ref{appB}):
\begin{equation}
    \mathcal{G}^{ab}(z,\bar{z})=\sum_{n,m\in\mathbb{Z}}\mathcal{G}^{ab}_{mn}z^{-m-1}\bar{z}^{-n-1},
    \label{eq:curlyGmode}
\end{equation}
with 
\begin{equation}
    \mathcal{G}^{ab}_{mn}=\frac{1}{(2\pi i)^2}\oint dz\oint d\bar{z} ~z^{m}\bar{z}^{n}\mathcal{G}^{ab}(z,\bar{z}).
\end{equation}
From the OPE of Eq. \eqref{eq:opecurlyGO}, we have
\[
\begin{split}
    \left[\mathcal{G}^{ab}_{mn},\mathcal{O}^c_{\Delta,\ell}(w,\bar{w})\right]&=\frac{1}{(2\pi i)^2}\oint dz\oint d\bar{z}~ z^{m}\bar{z}^{n}\left[\mathcal{G}^{ab}(z,\bar{z}),\mathcal{O}^c_{\Delta,\ell}(w,\bar{w})\right]\\&=\sum_{d,e}f^{abd}f^{dce}\frac{1}{(2\pi i)^2}\oint dz\oint d\bar{z} ~z^{m}\bar{z}^{n}\frac{1}{z-w}\frac{1}{\bar{z}-\bar{w}}\mathcal{O}^e_{\Delta,\ell}(w,\bar{w})\\&=w^{m}\bar{w}^{n}\sum_{d,e}f^{abd}f^{dce}\mathcal{O}^e_{\Delta,\ell}(w,\bar{w}).
\end{split}
\]
In a similar way, using the OPEs in Eq. \eqref{eq:opeGO}, \eqref{eq:opeGbarO} and the integral expression  \eqref{eq:modesint} for the modes $G^a_m$ and $\overline{G}^b_n$, we have 
\[
\begin{split}
\left[G^a_m,\mathcal{O}^c_{\Delta,\ell}(w,\bar{w})\right]=-w^{m}\sum_{d}f^{acd}\mathcal{O}^d_{\Delta,\ell}(w,\bar{w})\\\left[\overline{G}^a_n,\mathcal{O}^c_{\Delta,\ell}(w,\bar{w})\right]=-\bar{w}^{n}\sum_{d}f^{acd}\mathcal{O}^d_{\Delta,\ell}(w,\bar{w}).    
\end{split}
\]
Thus we have 
\[
\begin{split}
    \left[\left[G^a_m,\overline{G}^b_n\right],\mathcal{O}^c_{\Delta,\ell}(w,\bar{w})\right]&=\left[G^a_m,\left[\overline{G}^b_n,\mathcal{O}^c_{\Delta,\ell}(w,\bar{w})\right]\right]-\left[\overline{G}^b_n,\left[G^a_m,\mathcal{O}^c_{\Delta,\ell}(w,\bar{w})\right]\right]\\&=-\bar{w}^{n}\sum_df^{bcd}\left[G^a_m,\mathcal{O}^d_{\Delta,\ell}(w,\bar{w})\right]+w^{m}\sum_d f^{acd}\left[\overline{G}^b_n,\mathcal{O}^d_{\Delta,\ell}(w,\bar{w})\right]\\&=w^{m}\bar{w}^{n}\left[\sum_{d,e}f^{bcd}f^{ade}\mathcal{O}^e_{\Delta,\ell}(w,\bar{w})-\sum_{d,e}f^{acd}f^{bde}\mathcal{O}^e_{\Delta,\ell}(w,\bar{w})\right]\\&=w^{m}\bar{w}^{n}\sum_{d,e}f^{abd}f^{dce}\mathcal{O}^e_{\Delta,\ell}(w,\bar{w})\\&=\left[\mathcal{G}^{ab}_{mn},\mathcal{O}^c_{\Delta,\ell}(w,\bar{w})\right].
\end{split}
\]
The above expression further justifies the equality of the two operators as 
\begin{equation}
    \mathcal{G}^{ab}_{mn} =[G^{a}_m, \overline{G}^{b}_{n}]= G^{a}_m \overline{G}^{b}_{n}-\overline{G}^{b}_{n} G^{a}_{m}.
\label{eq:curlyGmnGnGn}
\end{equation}
Next we look for, 
\begin{equation}\label{newcr1}
\begin{split}
    \left[\mathcal{G}^{ab}_{m n}, L_{k}\right]&=\left[G^{a}_m \overline{G}^{b}_{n}-\overline{G}^{b}_{n} G^{a}_{m}, L_{k}\right]
    \\&=G^{a}_m\left[\overline{G}^{b}_{n}, L_{k}\right]+\left[G^{a}_{m}, L_{k}\right] \overline{G}^{b}_{n}-\overline{G}^{b}_{n}\left[G^{a}_{m}, L_{k}\right]-\left[\overline{G}^{b}_{n}, L_{k}\right] G^{a}_{m}\\
&=m G^{a}_{m+k} \overline{G}^{b}_{n}-m\overline{G}^{b}_{n} G^{a}_{m+k}\\
&= m \mathcal{G}^{ab}_{m+k, n}.
\end{split}
\end{equation}
Similarly,
\begin{equation}\label{newcr2}
    \quad \left[\mathcal{G}^{ab}_{m n}, \bar{L}_{k}\right]= n\mathcal{G}^{ab}_{m, n+k}
\end{equation}
Finally, to commute the commutator of $\mathcal{G}^{ab}_{mn}$ with $\mathcal{G}^{a'b'}_{kl}$, we compute its commutator with an arbitary primary operator $\mathcal{O}^c_{\Delta,\ell}(w,\bar{w})$. We have
\[
\begin{split}
    \left[\left[\mathcal{G}^{ab}_{mn},\mathcal{G}^{a'b'}_{kl}\right],\mathcal{O}^c_{\Delta,\ell}(w,\bar{w})\right]&=\left[\mathcal{G}^{ab}_{mn},\left[\mathcal{G}^{a'b'}_{kl},\mathcal{O}^c_{\Delta,\ell}(w,\bar{w})\right]\right]-\left[\mathcal{G}^{a'b'}_{kl},\left[\mathcal{G}^{ab}_{mn},\mathcal{O}^c_{\Delta,\ell}(w,\bar{w})\right]\right]\\&=w^{k}\bar{w}^{l}\sum_{d,e}f^{a'b'd}f^{dce}\left[\mathcal{G}^{ab}_{mn},\mathcal{O}^e_{\Delta,\ell}(w,\bar{w})\right]\\&\hspace{4cm}-w^{m}\bar{w}^{n}\sum_{d,e}f^{abd}f^{dce}\left[\mathcal{G}^{a'b'}_{kl},\mathcal{O}^e_{\Delta,\ell}(w,\bar{w})\right]\\&=w^{m+k}\bar{w}^{n+l}\left[\sum_{d,e}f^{a'b'd}f^{dce}\sum_{d',e'}f^{abd'}f^{d'ee'}\mathcal{O}^{e'}_{\Delta,\ell}(w,\bar{w})\right.\\&\left.\hspace{4cm}-\sum_{d,e}f^{abd}f^{dce}\sum_{d',e'}f^{a'b'd'}f^{d'ee'}\mathcal{O}^{e'}_{\Delta,\ell}(w,\bar{w})\right] \\&=\sum_{d,d'}f^{abd'}f^{a'b'd}\left(w^{m+k}\bar{w}^{n+l}\sum_{e,e'}f^{d'de}f^{ece'}\mathcal{O}^{e'}_{\Delta,\ell}(w,\bar{w})\right) \\
    &=\sum_{d,d'}f^{abd}f^{a'b'd}\left[\mathcal{G}^{d'd}_{m+k,n+l},\mathcal{O}^c_{\Delta,\ell}(w,\bar{w})\right].
    \end{split}
\]
where we have used the identity of structure constants for simplification. Hence we conclude that 
\begin{equation}
\left[\mathcal{G}^{ab}_{mn},\mathcal{G}^{a'b'}_{kl}\right]=\sum_{d,d'}f^{abd'}f^{a'b'd}\mathcal{G}^{d'd}_{m+k,n+l}.    
\end{equation}
Now collecting all the commutators we get the complete set of algebra as, 
\begin{equation}\label{eq:bmsalgeym}
\begin{split}
\begin{aligned}[c]
&{\left[L_{m}, L_{n}\right]=(m-n) L_{m+n}} \\
&{\left[\overline{L}_{m}, \overline{L}_{n}\right]=(m-n) \overline{L}_{m+n}}\\
&{\left[L_{n}, P_{k l}\right]=\left(\frac{1}{2} n-k\right) P_{n+k, l}} \\
&{\left[\bar{L}_{n}, P_{k l}\right]=\left(\frac{1}{2} n-l\right) P_{k, n+l}}\\
\end{aligned}
\qquad \qquad
\begin{aligned}[c]
&\left[\mathcal{G}^{ab}_{m n}, P_{k,l}\right]= 0\\
&\left[\mathcal{G}^{ab}_{m n}, L_{k}\right]= m \mathcal{G}^{ab}_{m+k, n}\\
&\left[\mathcal{G}^{ab}_{m n}, \overline{L}_{k}\right]= n\mathcal{G}^{ab}_{m, n+k},\\&\left[\mathcal{G}^{ab}_{mn},\mathcal{G}^{a'b'}_{kl}\right]=\sum_{d,d'}f^{abd'}f^{a'b'd}\mathcal{G}^{d'd}_{m+k,n+l}.
\end{aligned}
\end{split}
\end{equation}
The above algebra \eqref{eq:bmsalgeym} is our  $\mathfrak{bms}_4$ with an $\mathrm{u}(N)$ extension \cite{Fotopoulos:2019vac}. This is the infinite-dimensional symmetry algebra of Einstein-Yang-Mills theory on $\mathcal{CS}^2$ and can be compared to the asymptotic symmetry algebra of the same as given in \cite{Barnich:2013sxa}. On comparing with \cite{Barnich:2013sxa}, we see that our method gives exactly the same algebra as theirs, but in a different basis. In particular, we see that the generator $j_{i}^{m,n}$ of \cite{Barnich:2013sxa} is related to our generator $\mathcal{G}^{ab}_{mn}$ as follows:
\[
\mathcal{G}^{ab}_{mn}=\sum_{c}f^{abc}j_{c}^{m,n}.
\]
Thus we see that the asymptotic symmetry algebra $\mathfrak{e}\textbf{y}\mathfrak{mbms}_4$ of EYM-theory has the structure of the semi-direct sum of superrotations with the direct sum of the abelian alegbra of supertranslations and nonabelian $\mathrm{u}(N)$-gauge transformations represented by $\mathcal{G}^{ab}_{mn}$: 
\begin{equation*}
    \mathfrak{e}\textbf{y}\mathfrak{mbms}_4=\text{Superrotations}\uplus\left[\text{Supertranslations}\oplus \mathrm{u}(N)\text{-gauge transformations}\right].
\end{equation*}
In the above formalism by introducing the conformal operator for $\mathrm{u}(N)$  gauge bosons, we have generated the symmetry algebra of the corresponding theory. In the next section we follow the same procedure to find the asymptotic symmetry algebra of Einstein-Maxwell theory by introducing the conformal operator of a single gauge boson.

\section{Asymptotic Symmetry of EM Theory}\label{asyEM}
\noindent
In this section we look for the asymptotic symmetry for Einstein-Maxwell theory, using the OPEs of the corresponding $\mathcal{CS}^2$ amplitudes. 
Following the same prescription as of the last section
we define the current corresponding to the conformal operator of the photon $A^{\mu}$ which is supposed to generate the $\mathrm{u}(1)$-gauge transformations in a similar way:
\begin{equation}
 \begin{aligned}
&G(z) = \frac{1}{2 \pi} \int d^{2} z^{\prime} \frac{1}{\left(z-z^{\prime}\right)^{2}} \mathcal{O}_{1,-1}(z, \bar{z}) \\
&\overline{G}(\bar{z}) =\frac{1}{2 \pi} \int d^{2} z^{\prime} \frac{1}{\left(\bar{z}-\bar{z}^{\prime}\right)^{2}} \mathcal{O}_{1,+1}(z, \bar{z})
\end{aligned}
\label{eq:GGbardef}   
\end{equation}

We now calculate the OPE of $G$ and $\overline{G}$ with spin 1 primary operators $\mathcal{O}_{\Delta,\ell}$. For $M$ such primary operators, we have by Eq. \eqref{eq:Ocorscatamprel}
\[
\left\langle G(z) \prod_{n=1}^{M} \mathcal{O}_{\Delta_{n}, \ell_{n}}\left(z_{n}, \bar{z}_{n}\right)\right\rangle=\frac{1}{2 \pi} \int d^{2} z_{0} \frac{1}{\left(z-z_0\right)^{2}}\left\langle\mathcal{O}_{1,-1}(z_0,\bar{z}_0)\prod_{n=1}^{M} \mathcal{O}_{\Delta_{n}, \ell_{n}}\left(z_{n}, \bar{z}_{n}\right)\right\rangle.
\]

When the gauge group is $\mathrm{u}(1)$, the soft theorem does not involve the gauge group factors. In particular for $M$ bosons interacting in EM-theory, the correlator of the corresponding conformal operators $\mathcal{O}_{\Delta_i,\ell_i}$ is related to the scattering amplitude $\mathcal{A}_{\ell_1,\dots,\ell_M}$ in Mellin space as follows (cf. \cite[Eq. 3.30]{Fan:2019emx}):
\begin{equation}
    \left\langle\prod_{n=1}^M\mathcal{O}_{\Delta_n,\ell_n}(z_n,\bar{z}_n)\right\rangle=\sum_{\sigma\in S_{M-1}}\mathcal{A}^{\sigma}_{\ell_1,\dots,\ell_M}(z_n,\bar{z}_n),
\end{equation}
where $\mathcal{A}_{\ell_1,\dots,\ell_M}^{\sigma}$ is the partial amplitude corresponding to the permutation of the $M-1$ external legs fixing the first leg.
The soft limit of \cite[Eq. (3.32)]{Fan:2019emx} transforms to  
\begin{equation}
\begin{split}
    \Bigg{\langle}\mathcal{O}_{1,-1}(z_0,\bar{z}_0)\prod_{n=1}^{M} &\mathcal{O}_{\Delta_{n}, \ell_{n}}\left(z_{n}, \bar{z}_{n}\right)\Bigg{\rangle}\\&=\sum_{i=1}^M\frac{1}{\bar{z}_0-\bar{z}_i}\left\langle\mathcal{O}_{\Delta_{1} ,\ell_{1}}\left(z_{1}, \bar{z}_{1}\right) \ldots \mathcal{O}_{\Delta_{i}, \ell_{i}}\left(z_{i}, \bar{z}_{i}\right) \ldots \mathcal{O}_{\Delta_{n}, 
    \ell_{n}}\left(z_{n}, \bar{z}_{n}\right)\right\rangle
\end{split}
\label{eq:YMsoftlim}
\end{equation}
Using this and following the calculations on previous sections, we easily see that 
\begin{equation}
    \begin{split}
    &\left\langle G(z) \prod_{n=1}^{M} \mathcal{O}_{\Delta_{n}, \ell_{n}}\left(z_{n}, \bar{z}_{n}\right)\right\rangle=-\sum_{i=1}^M\frac{1}{z-z_i}\left\langle\mathcal{O}_{\Delta_{1} ,\ell_{1}}\left(z_{1}, \bar{z}_{1}\right) \ldots \mathcal{O}_{\Delta_{i}, \ell_{i}}\left(z_{i}, \bar{z}_{i}\right) \ldots \mathcal{O}_{\Delta_{n}, 
    \ell_{n}}\left(z_{n}, \bar{z}_{n}\right)\right\rangle\\&\left\langle \overline{G}(\bar{z}) \prod_{n=1}^{M} \mathcal{O}_{\Delta_{n}, \ell_{n}}\left(z_{n}, \bar{z}_{n}\right)\right\rangle=-\sum_{i=1}^M\frac{1}{\bar{z}-\bar{z}_i}\left\langle\mathcal{O}_{\Delta_{1} ,\ell_{1}}\left(z_{1}, \bar{z}_{1}\right) \ldots \mathcal{O}_{\Delta_{i}, \ell_{i}}\left(z_{i}, \bar{z}_{i}\right) \ldots \mathcal{O}_{\Delta_{n}, 
    \ell_{n}}\left(z_{n}, \bar{z}_{n}\right)\right\rangle
\end{split}
\end{equation}
This gives the OPE
\begin{equation}\label{opeGo}
\begin{split}
    &G(z)\mathcal{O}_{\Delta,\ell}(w,\bar{w})=\frac{1}{w-z}\mathcal{O}_{\Delta ,\ell}(w,\bar{w})\\&\overline{G}(\bar{z})\mathcal{O}_{\Delta,\ell}(w,\bar{w})=\frac{1}{\bar{w}-\bar{z}}\mathcal{O}_{\Delta ,\ell}(w,\bar{w}).
\end{split}
\end{equation}
The OPEs of Sections \ref{sec:GG}, \ref{sec:PG} and \ref{sec:TG} remain the same without the gauge indices. Finally, we see that due to the absence of gauge factors $f^{abc}$, the OPE of the new current 
\[
\mathcal{G}(z,\bar{z})=~:G(z)\overline{G}(\bar{z})-\overline{G}(\bar{z})G(z):
\]
is regular:
\begin{equation}
    \left\langle\mathcal{G}(z,\bar{z}) \prod_{n=1}^{M} \mathcal{O}_{\Delta_{n}, \ell_n }\left(z_{n}, \bar{z}_{n}\right)\right\rangle\sim\text{regular.}
\end{equation}
To find out the algebra, we Laurent expand the currents:
\begin{equation}
    G(z)=\sum_{m \in \mathbb{Z}} \frac{G_{m}}{z^{m+1}},  \quad \overline{G}(\bar{z})=\sum_{n \in \mathbb{Z}} \frac{\overline{G}_{n}}{\bar{z}^{n+1}}
\end{equation}
where,
\begin{equation}
    \begin{split}
    G_{m}=\frac{1}{2 \pi i} \oint d z \; z^{m} G(z), \quad \overline{G}_{n}=\frac{1}{2 \pi i} \oint d \bar{z}\;  \bar{z}^{n} \overline{G}(\bar{z}) .
\end{split}
\label{eq:modesint}
\end{equation}
The algebra remains the same as in Eq. \eqref{eq:bmsalgeym} except that now the commutator 
\[
[G_m,\overline{G}_n]=0,
\]
as can easily be verified using the OPE in Eq. \eqref{opeGo} and following the same method as in the proof of Eq. \eqref{eq:curlyGmnGnGn}. We collect the algebra here for completeness:
\begin{equation}\label{eq:bmsalgem}
\begin{split}
&\left[G_{m}, P_{k, l}\right]=\left[\overline{G}_{m}, P_{k, l}\right]=0\\
&\left[L_{m}, \overline{G}_{n}\right]=\left[\overline{L}_{m}, G_{n}\right]=0\\
&\left[G_m,G_n\right]=\left[\overline{G}_m,\overline{G}_n\right]=\left[G_m,\overline{G}_n\right]=0\\
&\left[G_m, L_{n}\right]=m G_{m+n}\\
&\left[\overline{G}_{m}, \overline{L}_{n}\right]=m \overline{G}_{m+n}.
\end{split}
\end{equation}
From the algebra above and the algebra of supertranslations and superrotations, we conclude that the extended $\mathfrak{bms}_4$ algebra of EM-theory has the structure of semidirect sum of superrotations with the direct sum of the abelian algebras of supertranslations and $\mathrm{u}(1)$-gauge transformations \cite{ Henneaux:2018hdj}:
\begin{equation*}
    \mathfrak{embms}_4=\text{Superrotations}\uplus \left[\text{Supertranslations}\oplus \mathrm{u}(1)\text{-gauge transformations}\right].
\end{equation*}
We remark that the asymptotic algebra calculated using the methods of \cite{Henneaux:2018hdj} only gives the global Lorentz transformations although we have recovered the complete local superrotations algebra using our methods. 
\section{Conclusions and Open Problems}
  In this paper, we have used the CCFT technique to compute the asymptotic symmetry algebra of EYM and EM theories in four spacetime dimensions. We defined the celestial conformal operators corresponding to the symmetry currents of graviton and gauge boson fields of the bulk theory and computed their OPEs using the well known soft and collinear limits \cite{Taylor:2017sph}. The resultant correlators, in their singular (OPE) limit gives the bulk scattering amplitudes where one or more of the incoming particles are soft. Using the standard methods of 2D CFT we computed the symmetry algebra of modes of these current operators, that finally gives us the asymptotic symmetry algebra of the bulk theory.
  The asymptotic symmetry algebra of the EM theory is obtained from that of the EYM theory, by setting the non-abelian structure constants to zero value. These give us the extended $\mathfrak{bms}_4$ symmetry algebra, in presence of non-abelian and abelian spin 1 current. In the case of EYM theory, due to its non-abelian structure, a composite current conformal operator  playes an important role. Our results match with the known results in literature \cite{Barnich:2013sxa, Henneaux:2018hdj} upto a generator redefinition. This bolsters the proposal of Taylor \textit{et. al.} \cite{Schreiber:2017jsr, Barnich:2013sxa, Fotopoulos:2019vac, Adamo:2019ipt, Fan:2019emx, Fotopoulos:2020bqj} of using celestial CFT correlators to compute asymptotic symmetry algebra of flat field theories.\\

This method of computing asymptotic symmetry algebra using CCFT is efficient and computationally easier as compared to the usual Killing vector approach. 
Recently the CCFT approach has been extended to calculate the  asymptotic symmetry algebra of $\mathcal{N}=1$ supergravity theory \cite{Fotopoulos:2020bqj}. The asymptotic symmetry algebra in this case is an infinite dimensional extension of the $\mathfrak{bms}_4$ algebra by the fermionic current modes. This work made use of supersymmetric ward identities to extract the soft and collinear limits of conformal currents associated to gravitons and gravitinos. It would be interesting to extend this approach to other four dimensional $\mathcal{N}>1$ supersymmetric field and gravity theories. In $\mathcal{N}>1$ supersymmetric theories, due to the presence of $R$-charges, it is expected that the resultant extension of the asymptotic algebra will be ``nontrivial". One prime ingredient of the CCFT approach is the soft and collinear limits of the (super)conformal operators corresponding to the particle excitations in the theory. Thus, the extension to $\mathcal{N}>1$ case requires the study of soft and collinear singularities in those theories. Such a study appeared in \cite{Jiang:2021xzy} for $\mathcal{N}=4$ Super Yang-Mills theory. It would be interesting to use their results for finding the asymptotic symmetry algebra for  $\mathcal{N}=4$ Super Yang-Mills theory. However, even more interesting scenario would be to study the extension of the BMS algebra for supergravity theories with $\mathcal{N}>1$ supersymmetry. One such candidate is the $\mathcal{N}=8$ supergravity which is related to $\mathcal{N}=4$ Super Yang-Mills theory via the double copy relations and hence an analysis of soft and collinear limits would be tractable in this case. We hope to report on it in near future.

Before we conclude this paper, let us briefly discuss the importance of the study of asymptotic symmetry algebra for higher $\mathcal{N}$ supergravity theories. Firstly, in the context of flat space holography\footnote{In the context of 3dimensional (super)gravity, the duals have been cobstructed in \cite{Barnich:2013yka, Barnich:2015sca, Barnich:2014cwa, Caroca:2018obf, Banerjee:2019epe, Banerjee:2021uxl}.}, they can be used to find a field theory dual of the bulk theories. Supersymmetry gives us more technical handle to address such questions. Secondly, it is conjectured \cite{Hawking:2016msc, Averin:2016ybl, Donnay:2015abr} that the BMS hairs are responsible for Black Hole entropy. The conjecture has also been shown to be only partially correct in \cite{Mirbabayi:2016axw, Bousso:2017dny}, where the authors showed that the BMS hair can only partially  incorporate some part of the Black Hole entropy. In the context of a class of black holes, namely extremal black holes in (super)string theories, the microscopic counting of black hole states are known in great details. It would be interesting to find how much of this entropy is captured by the (super)$\mathfrak{bms}_4$ hairs. This project remains as one of our prime goal to study in future.

\newpage

\section*{Acknowledgement}
We would like to thank Arpita Mitra, Debanshu Mukherjee, H. R. Safari and Shahin Sheikh-Jabbari for helpful discussions. R.K.S would like to acknowledge the hospitality of IISER Bhopal where part of this work was done. The work is partially supported by SERB ECR grant, Govt. of India. Finally we acknowledge the support of people of India towards fundamental research. 

\appendix
\section{Certain Integrals}
We show that certain integral vanish as a result of global conformal invariance of celestial correlators. These results are used in the computations presented in the main draft.
\subsection{}\label{appA1}
We have,
\[
\begin{split}
     \left\langle G^a(z) G^b(w) \prod_{n=2}^{N}\mathcal{O}_{\Delta_{n}, \ell_{n}}\left(z_{n}, \bar{z}_{n}\right)\right\rangle &=- \frac{1}{2\pi}\sum_{i=1}^M\sum_{c}f^{ab_ic}\int d^{2} z_{1} \frac{1}{\left(w-z_{1}\right)^{2}}\frac{1}{z-z_i}\\
    & \times \left\langle\mathcal{O}_{1 ,-1}^{b_{1}=b}\left(z_{1}, \bar{z}_{1}\right) \ldots \mathcal{O}_{\Delta_{i}, \ell_{i}}^{c}\left(z_{i}, \bar{z}_{i}\right) \ldots \mathcal{O}_{\Delta_{n}, 
    \ell_{n}}^{b_{n}}\left(z_{n}, \bar{z}_{n}\right)\right\rangle.
\end{split}
\]
\\
Now using the global conformal invariance of $\left\langle\mathcal{O}_{1 ,-1}^{b_{1}=b}\left(z_{1}, \bar{z}_{1}\right) \ldots \mathcal{O}_{\Delta_{i}, \ell_{i}}^{c}\left(z_{i}, \bar{z}_{i}\right) \ldots \mathcal{O}_{\Delta_{n}, 
    \ell_{n}}^{b_{n}}\left(z_{n}, \bar{z}_{n}\right)\right\rangle$, we will show that the first integral vanishes and the only nontrivial terms which exist are the regular ones. \\
    
Putting $W = w-z_1$, $Z= z-z_1$ we have,
\[
\begin{split}
    \int d^{2} z_{1} \; &\frac{1}{W^{2} Z} \; \left\langle\mathcal{O}_{1 ,-1}^{b_{1}=b}\left(z_{1}, \bar{z}_{1}\right) \ldots \mathcal{O}_{\Delta_{i}, \ell_{i}}^{c}\left(z_{i}, \bar{z}_{i}\right) \ldots \mathcal{O}_{\Delta_{n}, 
    \ell_{n}}^{b_{n}}\left(z_{n}, \bar{z}_{n}\right)\right\rangle\\
   & = -\partial_{w} \int d^{2} z_{1} \; \frac{1}{W Z} \;\left\langle\mathcal{O}_{1 ,-1}^{b_{1}=b}\left(z_{1}, \bar{z}_{1}\right) \ldots \mathcal{O}_{\Delta_{i}, \ell_{i}}^{c}\left(z_{i}, \bar{z}_{i}\right) \ldots \mathcal{O}_{\Delta_{n}, 
    \ell_{n}}^{b_{n}}\left(z_{n}, \bar{z}_{n}\right)\right\rangle.
\end{split}
\]
We use the following identity:
\begin{equation*}
    \begin{split}
    \begin{aligned}
    \frac{1}{WZ}&=\frac{1}{z-w}\left(\frac{1}{w-z_{1}}-\frac{1}{z-z_{1}}\right) =\frac{1}{z-w}\left(\frac{1}{W}-\frac{1}{Z}\right)
    \end{aligned}
\end{split}
\label{eq:idWZ=W-Z}
\end{equation*}
We get,
\[
\begin{split}
  \int d^{2} z_{1} \; &\frac{1}{W Z} \;\left\langle\mathcal{O}_{1 ,-1}^{b_{1}=b}\left(z_{1}, \bar{z}_{1}\right) \ldots \mathcal{O}_{\Delta_{i}, \ell_{i}}^{c}\left(z_{i}, \bar{z}_{i}\right) \ldots \mathcal{O}_{\Delta_{n}, 
    \ell_{n}}^{b_{n}}\left(z_{n}, \bar{z}_{n}\right)\right\rangle\\
    &=  \frac{1}{z-w} \int d^{2} z_{1}\; \left(\frac{1}{W}-\frac{1}{Z}\right) \; \left\langle\mathcal{O}_{1 ,-1}^{b_{1}=b}\left(z_{1}, \bar{z}_{1}\right) \ldots \mathcal{O}_{\Delta_{i}, \ell_{i}}^{c}\left(z_{i}, \bar{z}_{i}\right) \ldots \mathcal{O}_{\Delta_{n}, 
    \ell_{n}}^{b_{n}}\left(z_{n}, \bar{z}_{n}\right)\right\rangle.
\end{split}
\]
We change $z_1 \to z-w+z_1$ in second term of this integral:
\[
\begin{split}
 \int d^{2} z_{1}\; \frac{1}{z-z_1} \; &\left\langle\mathcal{O}_{1 ,-1}^{b_{1}=b}\left(z_{1}, \bar{z}_{1}\right) \ldots \mathcal{O}_{\Delta_{i}, \ell_{i}}^{c}\left(z_{i}, \bar{z}_{i}\right) \ldots \mathcal{O}_{\Delta_{n}, 
    \ell_{n}}^{b_{n}}\left(z_{n}, \bar{z}_{n}\right)\right\rangle \longrightarrow \\
 &\int d^{2} z_{1}\; \frac{1}{w-z_1} \; \left\langle\mathcal{O}_{1 ,-1}^{b_{1}=b}\left(z_{1}, \bar{z}_{1}\right) \ldots \mathcal{O}_{\Delta_{i}, \ell_{i}}^{c}\left(z_{i}, \bar{z}_{i}\right) \ldots \mathcal{O}_{\Delta_{n}, 
    \ell_{n}}^{b_{n}}\left(z_{n}, \bar{z}_{n}\right)\right\rangle
\end{split}
\]
By global conformal invariance of $\left\langle\mathcal{O}_{1 ,-1}^{b_{1}=b}\left(z_{1}, \bar{z}_{1}\right) \ldots \mathcal{O}_{\Delta_{i}, \ell_{i}}^{c}\left(z_{i}, \bar{z}_{i}\right) \ldots \mathcal{O}_{\Delta_{n}, 
    \ell_{n}}^{b_{n}}\left(z_{n}, \bar{z}_{n}\right)\right\rangle$ under $\begin{pmatrix}  1 & z-w \\
    0 & 1
    \end{pmatrix} \in \mathrm{SL}(2,\mathbb{C})$, we get (cf. \cite[Eq. (3.2)]{Fan:2019emx}), 
\[
\begin{split}
    \bigg\langle\mathcal{O}_{1 ,-1}^{b_{1}=b}(z-w+z_1,\; &\bar{z}-\bar{w}+\bar{z}_{1}) \ldots \mathcal{O}_{\Delta_{i}, \ell_{i}}^{c}(z_{i}, \bar{z}_{i}) \ldots \mathcal{O}_{\Delta_{n}, \ell_{n}}^{b_{n}}(z_{n},\bar{z}_{n})\bigg\rangle \\
    &= \bigg\langle\mathcal{O}_{1 ,-1}^{b_{1}=b}\left(z_{1}, \bar{z}_{1}\right) \ldots \mathcal{O}_{\Delta_{i}, \ell_{i}}^{c}\left(z_{i}, \bar{z}_{i}\right) \ldots \mathcal{O}_{\Delta_{n}, \ell_{n}}^{b_{n}}\left(z_{n}, \bar{z}_{n}\right)\bigg\rangle,
\end{split}
\]
which implies 
\[
\begin{split}
    \int d^{2} z_{1} \; \frac{1}{W Z} \;\left\langle\mathcal{O}_{1 ,-1}^{b_{1}=b}\left(z_{1}, \bar{z}_{1}\right) \ldots \mathcal{O}_{\Delta_{i}, \ell_{i}}^{c}\left(z_{i}, \bar{z}_{i}\right) \ldots \mathcal{O}_{\Delta_{n}, \ell_{n}}^{b_{n}}\left(z_{n}, \bar{z}_{n}\right)\right\rangle =0
\end{split}
\]
Thus we get 
\[
\begin{split}
    \int d^{2} z_{1} \; \frac{1}{W^{2} Z} \; \left\langle\mathcal{O}_{1 ,-1}^{b_{1}=b}\left(z_{1}, \bar{z}_{1}\right) \ldots \mathcal{O}_{\Delta_{i}, \ell_{i}}^{c}\left(z_{i}, \bar{z}_{i}\right) \ldots \mathcal{O}_{\Delta_{n}, \ell_{n}}^{b_{n}}\left(z_{n}, \bar{z}_{n}\right)\right\rangle=-\partial_{w} (0) =0
\end{split}
\]
\subsection{}\label{appA2}
We show that 
\[
\frac{1}{2 \pi} \int d^{2} z_{1}\left(\frac{1}{z-w}\frac{1}{WZ}-\frac{1}{WZ^2}\right)\left\langle \mathcal{O}^{a}_{1,-1}\left(z_1, \bar{z}_{1}\right) \prod_{n=2}^{M} \mathcal{O}^{b_n}_{\Delta_{n}, \ell_n }\left(z_{n}, \bar{z}_{n}\right)\right\rangle=0.
\]
It suffices to prove that 
\[
\frac{1}{2 \pi} \int d^{2} z_{1}\frac{1}{WZ}\left\langle \mathcal{O}^{a}_{1,-1}\left(z_1, \bar{z}_{1}\right) \prod_{n=2}^{M} \mathcal{O}^{b_n}_{\Delta_{n}, \ell_n }\left(z_{n}, \bar{z}_{n}\right)\right\rangle=0,
\]
since the second term is given by 
\[
\begin{split}
\frac{1}{2 \pi} \int d^{2} z_{1}\frac{1}{WZ^2}\Bigg\langle \mathcal{O}^{a}_{1,-1}\left(z_1, \bar{z}_{1}\right) &\prod_{n=2}^{M} \mathcal{O}^{b_n}_{\Delta_{n}, \ell_n }\left(z_{n}, \bar{z}_{n}\right)\Bigg\rangle\\&=-\frac{1}{2 \pi} \partial_z\left[\int d^{2} z_{1}\frac{1}{WZ}\left\langle \mathcal{O}^{a}_{1,-1}\left(z_1, \bar{z}_{1}\right) \prod_{n=2}^{M} \mathcal{O}^{b_n}_{\Delta_{n}, \ell_n }\left(z_{n}, \bar{z}_{n}\right)\right\rangle\right]\\
&=0
\end{split}
\]
Using the global conformal invariance of the correlator as in \ref{appA1}, the proof is complete.
\section{Conformal Dimension of $\mathcal{G}^{ab}(z,\bar{z})$}\label{appB}
In this appendix, we explicitly compute the OPE of $\mathcal{G}^{ab}(z,\bar{z})$ with $T(z)$ and $\overline{T}(\bar{z})$ and conclude that the conformal dimension of $\mathcal{G}^{ab}(z,\bar{z})$ is $(h,\bar{h})=(1,1)$. This justifies the mode expansion of $\mathcal{G}^{ab}(z,\bar{z})$ in Eq. \eqref{eq:curlyGmode}. We have 
\[
\begin{split}
\left\langle T(z)\mathcal{G}^{ab}(w,\bar{w})\right\rangle&=\left\langle T(z)~:G^a(w)\overline{G}^b(\bar{w}):~\right\rangle-\left\langle T(z)~:\overline{G}^b(\bar{w})G^a(w):~\right\rangle
\end{split}
\]
which is the difference of two three point correlation functions. We can use generalised Wick's theorem (see \cite[Appendix 6.B]{di1996conformal}) to simplify this. Since the normal ordering `$:\ :$' removes the singular terms from the OPE, we can use the expression \cite[Appendix B, Eq. (6.206)]{di1996conformal}. Then we have 
\[
\begin{split}
  &\left\langle T(z)~:G^a(w)\overline{G}^b(\bar{w}):~\right\rangle\\
  &=\frac{1}{2\pi i}\oint\frac{dx}{x-w} \left[T\overbracket{(z)~:G^a(}x)\overline{G}^b(\bar{w}):+ :~G^a(x) T\overbracket{(z)~\overline{G}^b(}\bar{w}):\right]\\&=\frac{1}{2\pi i}\oint\frac{dx}{x-w} \Bigg[\frac{1}{(z-x)^2}:G^a(x)\overline{G}^b(\bar{w}):+\frac{1}{z-x}~:\partial_x G^a(x)\overline{G}^b(\bar{w}):~+~\text{regular.}\Bigg]\\&=\frac{1}{(z-w)^2}:G^a(w)\overline{G}^b(\bar{w}):+\frac{1}{z-w}:\partial_w G^a(w)\overline{G}^b(\bar{w}):,
\end{split}
\] 
where we used the OPE of Eq. \eqref{eq:opeTG} and Eq. \eqref{eq:TGbar}.
Similarly
\[
\begin{split}
&\left\langle T(z)~:\overline{G}^b(\bar{w})G^a(w):~\right\rangle\\
&=\frac{1}{2\pi i}\oint \frac{dx}{x-w} \left[T\overbracket{(z)~:\overline{G}^b(}\bar{w})G^a(x):+ :\overline{G}^b(\bar{w}) T\overbracket{(z)~G^a(}x):\right]\\&=\frac{1}{2\pi i}\oint\frac{dx}{x-w} \left[\text{regular.}~+~\frac{1}{(z-x)^2}:\overline{G}^b(\bar{w})G^a(x):+\frac{1}{z-x}~:\overline{G}^b(\bar{w})\partial_x G^a(x):\right]\\&=\frac{1}{(z-w)^2}:\overline{G}^b(\bar{w})G^a(w):+\frac{1}{z-w}:\overline{G}^b(\bar{w})\partial_w G^a(w):,
\end{split}
\]
where we again used the OPE of Eq. \eqref{eq:opeTG}.
Thus we get 
\[
\begin{split}
    \left\langle T(z)\mathcal{G}^{ab}(w,\bar{w})\right\rangle&=\frac{1}{(z-w)^2}:\left[G^a(w),\overline{G}^b(\bar{w})\right]:+\frac{1}{z-w}\partial_w:\left[G^a(w),\overline{G}^b(\bar{w})\right]:\\&=\frac{1}{(z-w)^2}\mathcal{G}^{ab}(w,\bar{w})+\frac{1}{z-w}\partial_w\mathcal{G}^{ab}(w,\bar{w}).
\end{split}
\]
Similarly we can compute the OPE of $\mathcal{G}^{ab}(z,\bar{z})$ with $\overline{T}(\bar{z})$. We get 
\[
\left\langle \overline{T}(\bar{z})\mathcal{G}^{ab}(w,\bar{w})\right\rangle=\frac{1}{(\bar{z}-\bar{w})^2}\mathcal{G}^{ab}(w,\bar{w})+\frac{1}{\bar{z}-\bar{w}}\partial_{\bar{w}}\mathcal{G}^{ab}(w,\bar{w}).
\]

\bibliography{bms.bib}

\end{document}